\documentclass[twocolumn]{aastex631}

\usepackage{amsmath}

\newcommand\uJy{$\mu$Jy}
\newcommand\magphysphotz{\texttt{MAGPHYS+photo}-$z$}
\newcommand\zpdf{$z_{\rm PDF}$}
\newcommand\qir{$q_{\rm IR}$}

\shorttitle{OIR-dark DSFGs in MORA}
\shortauthors{Manning et al.}

\begin{document}

\title{Characterization of Two 2\,mm-detected Optically-Obscured Dusty Star-Forming Galaxies}

\correspondingauthor{Sinclaire M. Manning}
\email{smanning.astro@gmail.com}

\author[0000-0003-0415-0121]{Sinclaire M. Manning}
\altaffiliation{NASA Hubble Fellow}
\affiliation{Department of Astronomy, The University of Texas at Austin, 2515 Speedway Boulevard Stop C1400, Austin, TX 78712, USA}
\affiliation{Department of Astronomy, University of Massachusetts, Amherst, MA 01003 USA}
\author[0000-0002-0930-6466]{Caitlin M. Casey}
\affiliation{Department of Astronomy, The University of Texas at Austin, 2515 Speedway Boulevard Stop C1400, Austin, TX 78712, USA}
\author[0000-0002-7051-1100]{Jorge A. Zavala}
\affiliation{National Astronomical Observatory of Japan, 2-21-1 Osawa, Mitaka, Tokyo 181-8588, Japan}
\affiliation{Department of Astronomy, The University of Texas at Austin, 2515 Speedway Boulevard Stop C1400, Austin, TX 78712, USA}
\author[0000-0002-4872-2294]{Georgios E. Magdis}
\affiliation{Cosmic Dawn Center (DAWN), Jagtvej 128, DK2200 Copenhagen N, Denmark}
\affiliation{Niels Bohr Institute, University of Copenhagen, Lyngbyvej 2, DK2100 Copenhagen \O, Denmark}
\affiliation{DTU-Space, Technical University of Denmark, Elektrovej 327, DK2800 Kgs. Lyngby, Denmark}
\author[0000-0003-3627-7485]{Patrick M. Drew}
\affiliation{Department of Astronomy, The University of Texas at Austin, 2515 Speedway Boulevard Stop C1400, Austin, TX 78712, USA}
\author[0000-0002-6184-9097]{Jaclyn B. Champagne}
\affiliation{Department of Astronomy, The University of Texas at Austin, 2515 Speedway Boulevard Stop C1400, Austin, TX 78712, USA}
\author[0000-0002-6290-3198]{Manuel Aravena}
\affiliation{N\'ucleo de Astronom\'ia, Facultad de Ingenier\'ia y Ciencias, Universidad Diego Portales, Av. Ej\'ercito 411, Santiago, Chile}
\author[0000-0002-3915-2015]{Matthieu Bethermin}
\affiliation{Aix Marseille Universit\'e, CNRS, LAM (Laboratoire d'Astrophysique de Marseille) UMR 7326, 13388, Marseille, France}
\author{David L. Clements}
\affiliation{Imperial College London, Blackett Laboratory, Prince Consort Road, London, SW7 2AZ, UK}
\author[0000-0001-8519-1130]{Steven L. Finkelstein}
\affiliation{Department of Astronomy, The University of Texas at Austin, 2515 Speedway Boulevard Stop C1400, Austin, TX 78712, USA}
\author[0000-0001-7201-5066]{Seiji Fujimoto}
\affiliation{Cosmic Dawn Center (DAWN), Jagtvej 128, DK2200 Copenhagen N, Denmark}
\affiliation{Niels Bohr Institute, University of Copenhagen, Lyngbyvej 2, DK2100 Copenhagen \O, Denmark}
\author[0000-0003-4073-3236]{Christopher C. Hayward}
\affiliation{Center for Computational Astrophysics, Flatiron Institute, 162 Fifth Avenue, New York, NY 10010, USA}
\author{Jacqueline A. Hodge}
\affiliation{Leiden Observatory, Leiden University, P.O. Box 9513, 2300 RA Leiden, the Netherlands}
\author[0000-0002-7303-4397]{Olivier Ilbert}
\affiliation{Aix Marseille Universit\'e, CNRS, LAM (Laboratoire d'Astrophysique de Marseille) UMR 7326, 13388, Marseille, France}
\author[0000-0001-9187-3605]{Jeyhan S. Kartaltepe}
\affiliation{School of Physics and Astronomy, Rochester Institute of Technology, Rochester, NY 14623, USA}
\author{Kirsten K. Knudsen}
\affiliation{Department of Space, Earth, and Environment, Chalmers University of Technology, Onsala Space Observatory, SE-439 92 Onsala, Sweden}
\author{Anton M. Koekemoer}
\affiliation{Space Telescope Science Institute, 3700 San Martin Dr., Baltimore, MD 21218, USA}
\author[0000-0003-2475-124X]{Allison W. S. Man}
\affiliation{Dunlap Institute for Astronomy \& Astrophysics, 50 St. George Street, Toronto, ON M5S 3H4, Canada}
\affiliation{Department of Physics \& Astronomy, University of British Columbia, 6224 Agricultural Road, Vancouver BC, V6T 1Z1, Canada}
\author{David B. Sanders}
\affiliation{Institute for Astronomy, 2680 Woodlawn Drive, University of Hawai'i, Honolulu, HI 96822, USA}
\author{Kartik Sheth}
\affiliation{NASA Headquarters, 300 E Street SW, Washington, DC 20546, USA}
\author[0000-0003-3256-5615]{Justin S. Spilker}
\affiliation{Department of Astronomy, The University of Texas at Austin, 2515 Speedway Boulevard Stop C1400, Austin, TX 78712, USA}
\author{Johannes Staguhn}
\affiliation{Department of Physics and Astronomy, Johns Hopkins University, 3400 North Charles Street, Baltimore, MD, 21218, USA}
\affiliation{Observational Cosmology Lab Code 665, NASA Goddard Space Flight Center, Greenbelt, MD, 20771, USA}
\author[0000-0003-4352-2063]{Margherita Talia}
\affiliation{Department of Physics and Astronomy ``Augusto Righi'' (DIFA), University of Bologna, Via Gobetti 93/2, I-40129, Bologna, Italy}
\affiliation{INAF - Osservatorio di Astrofisica e Scienza dello Spazio, Via Gobetti 93/3, I-40129, Bologna, Italy}
\author[0000-0001-7568-6412]{Ezequiel Treister}
\affiliation{Instituto de Astrof\'isica, Facultad de F\'isica, Pontificia Universidad Cat\'olica de Chile, Casilla 306, Santiago 22, Chile}
\author[0000-0001-7095-7543]{Min S. Yun}
\affiliation{Department of Astronomy, University of Massachusetts, Amherst, MA, 01003, USA }

\begin{abstract}
The 2\,mm Mapping Obscuration to Reionization with ALMA (MORA) Survey was designed to detect high redshift ($z\gtrsim4$), massive, dusty star-forming galaxies (DSFGs). Here we present two, likely high redshift sources, identified in the survey whose physical characteristics are consistent with a class of optical/near-infrared (OIR) invisible DSFGs found elsewhere in the literature. We first perform a rigorous analysis of all available photometric data to fit spectral energy distributions and estimate redshifts before deriving physical properties based on our findings. Our results suggest the two galaxies, called MORA-5 and MORA-9, represent two extremes of the ``OIR-dark" class of DSFGs. MORA-5 ($z_{\rm phot}=4.3^{+1.5}_{-1.3}$) is a significantly more active starburst with a star-formation rate of 830$^{+340}_{-190}$\,M$_\odot$\,yr$^{-1}$ compared to MORA-9 ($z_{\rm phot}=4.3^{+1.3}_{-1.0}$) whose star-formation rate is a modest 200$^{+250}_{-60}$\,M$_\odot$\,yr$^{-1}$. Based on the stellar masses (M$_{\star}\approx10^{10-11}$\,M$_\odot$), space density ($n\sim(5\pm2)\times10^{-6}$\,Mpc$^{-3}$, which incorporates two other spectroscopically confirmed OIR-dark DSFGs in the MORA sample at $z=4.6$ and $z=5.9$), and gas depletion timescales ($<1$\,Gyr) of these sources, we find evidence supporting the theory that OIR-dark DSFGs are the progenitors of recently discovered $3<z<4$ massive quiescent galaxies. 
\end{abstract}


\section{Introduction}
Our current understanding of star formation and galaxy evolution within the first two billion years after the Big Bang is severely limited by a lack of infrared (IR) constraints and associated sample incompleteness at $z>4$ (\citealt{Gruppioni2013MNRAS.432...23G}; \citealt*{Casey2014PhR...541...45C,MadauDickinson2014ARA&A}; \citealt{2018ApJ...862...77C}). Very recent studies \citep{Zavala2021ApJ...909..165Z,Casey2021arXiv211006930C} have worked to extend IR measurements up to $z\sim7$, but uncertainties are still dominant. The census of cosmic star formation out to the highest redshifts is also biased towards unobscured star formation tracers, relying on rest-frame ultraviolet (UV) continuum measurements to seek out Lyman-break selected galaxies (LBGs). While LBG-based studies have yielded the most robust results out to $z\sim6$ and beyond and have contributed heavily to analyses of the UV luminosity function (UVLF) \citep{ReddySteidel2009ApJ,Finkelstein2015ApJ...810...71F,Ono2018PASJ70S10O}, it is well understood that they miss both heavily and moderately dust-obscured star-formation by design \citep{Magnelli2012A&A...539A.155M,Gruppioni2013MNRAS.432...23G}. Therefore, the true abundance of dusty star-forming galaxies (DSFGs) in the early Universe remains unclear. Knowing the prevalence of dust-obscured star formation is particularly important at $z>4$ (within the first 1.5 billion years of the Universe) when cosmic time becomes a significant constraint on the physical processes which produce the dust, metals, and stars observed in galaxies. 

Much debate exists over the ubiquity of DSFGs at high redshifts and their contribution to the total cosmic star-formation within the first billion years after the Big Bang. The detection of these dust obscured systems require observations at far-IR and (sub)millimeter wavelengths which trace the dust re-processed UV/optical emission from young stars: an historically tricky task given the small area of deep pencil-beam observations and conversely the shallow depths reached by wider surveys. While some works propose DSFGs as the dominant source of star-formation at $z>4$ \citep{RowanRobinson2016MNRAS.461.1100R,Gruppioni2020A&A...643A...8G}, others find the contribution from DSFGs to be insignificant \citep{Koprowski2017MNRAS.471.4155K,Smith2017MNRAS.471.2453S}. In an attempt to reconcile these two disparate theories, we look to identify individual DSFGs and constrain their source density at $z>4$.

Models \citep{2018ApJ...862...78C,2018ApJ...862...77C} and observations \citep{Staguhn2014ApJ...790...77S} show that 2\,mm observations offer an effective strategy to detect high redshift dusty galaxies at $z\gtrsim4$ while simultaneously filtering out lower redshift DSFGs at $z\lesssim2.5$. The Mapping Obscuration to Reionization ALMA (MORA) Survey is the largest (184\,arcmin$^2$) 2\,mm mosaic mapped with the Atacama Large Millimeter/submillimeter Array (ALMA) whose aim is to efficiently select $z\gtrsim3-4$ DSFGs. For a full description of the MORA Survey, see \cite{Zavala2021ApJ...909..165Z} and \cite{Casey2021arXiv211006930C}. Twelve total sources were robustly detected ($>5\sigma$) in the MORA survey. Two of these sources were then identified as being of particular interest given their robust mm detections, but lack of optical/near-IR (OIR) counterparts. Systems with these characteristics have the potential to be the highest redshift galaxies in the sample or exceedingly obscured (or both). Additionally, recent works have unearthed a population of massive galaxies undetected in deep ($5\,\sigma$ point-source limit H-band\,$>27$\,mag) \textit{Hubble Space Telescope} observations (dubbed {\it HST}-dark or rest-frame {\it H}/{\it K}-band dropouts). As such, we focus on these two sources, MORA-5 and MORA-9, as they appear to be representative of this unique class of heavily obscured galaxies. ``OIR-dark'' is used as an identifier throughout this paper in reference to DSFGs with non-detections at ultraviolet, optical, and near-IR wavelengths shorter than 2.2\,\micron. 

In this paper we present the physical characteristics of two OIR-dark sources found in the 2\,mm MORA survey and discuss their potential to belong to the parent population of high-$z$ massive quiescent galaxies. Section \ref{sec:data_and_obs} presents the ALMA observations from MORA and archival searches as well as ancillary data from the COSMOS survey \citep{Scoville2007ApJS..172....1S}. Section \ref{sec:sed_fitting_techniques} describes the various spectral energy distribution (SED) fitting techniques used to fit the photometric data. Section \ref{sec:results} lists the physical properties derived from SED fitting. Section \ref{sec:discussion} discusses how these galaxies fit into the broader DSFG population and compare to known OIR-dark sources in the literature. Finally, we assess the potential evolutionary pathway of OIR-dark DSFGs as the progenitors of massive quiescent galaxies at early times. Throughout this paper we assume a Planck cosmology with $H_0=67.8$\,km\,s$^{-1}$\,Mpc$^{-1}$ and $\Omega_{\rm m}=0.308$ \citep{Planck2016A&A...594A..13P} and a Kroupa initial mass function \citep{Kroupa2003ApJ...598.1076K} when referring to SFRs.

\section{Data \& Observations}
\label{sec:data_and_obs}
In total, thirteen robust sources ($>5\,\sigma$) were detected in the 2\,mm mosaic. We direct the reader to the accompanying MORA papers for a more thorough overview of the survey and complete discussions on 2\,mm number counts \citep{Zavala2021ApJ...909..165Z} and source characterization \citep{Casey2021arXiv211006930C}. What follows is an explanation of the ALMA data reduction and imaging, including available archival data, as well as the existing ancillary data for MORA-5 and MORA-9, which are the focus of this paper. All available photometry is compiled in Table \ref{tab:photmetry}. 

\begin{figure*}
    \centering
    \includegraphics[width=\textwidth]{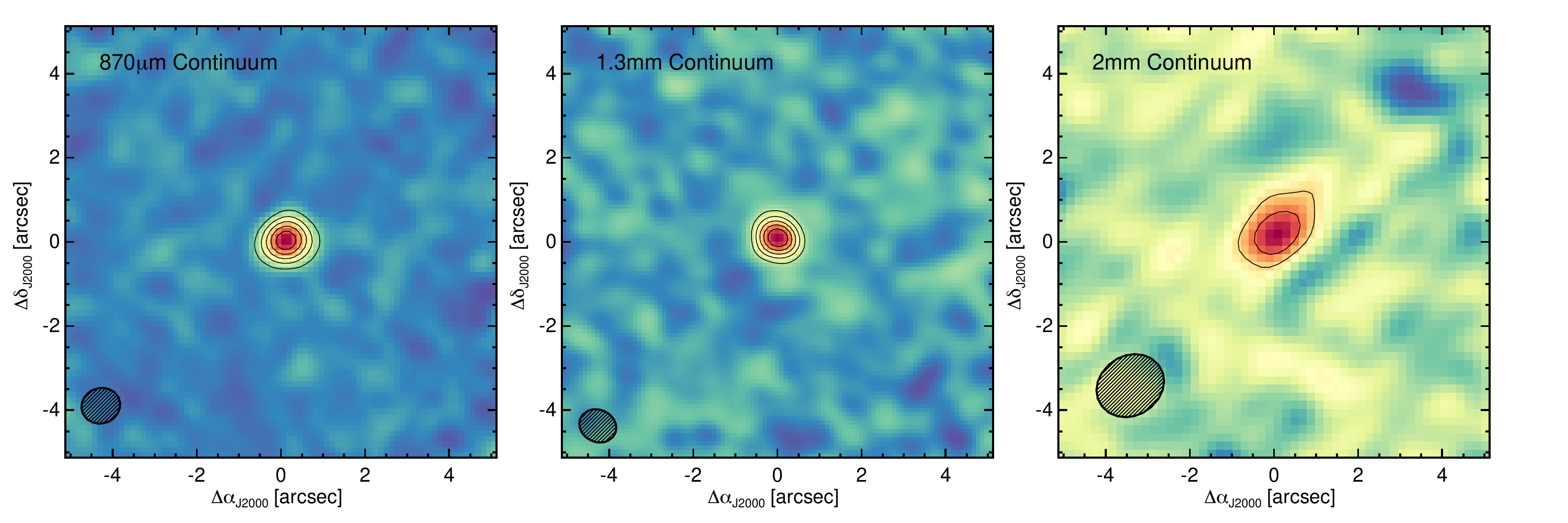}
    \caption{$10\arcsec\times10\arcsec$ cutouts of 870\,\micron, 1.3\,mm, and 2\,mm continuum imaging for MORA-5 using natural weighting (robust$=2$). Contours in the 870\,\micron\ image show 5, 10, 15, 20, and 25\,$\sigma$, in contrast to the 1.3\,mm and 2\,mm images which depict 3, 5, 7, 9, and 11\,$\sigma$.}
    \label{fig:source5_maps}
\end{figure*}

\begin{figure*}
    \centering
    \includegraphics[width=\textwidth]{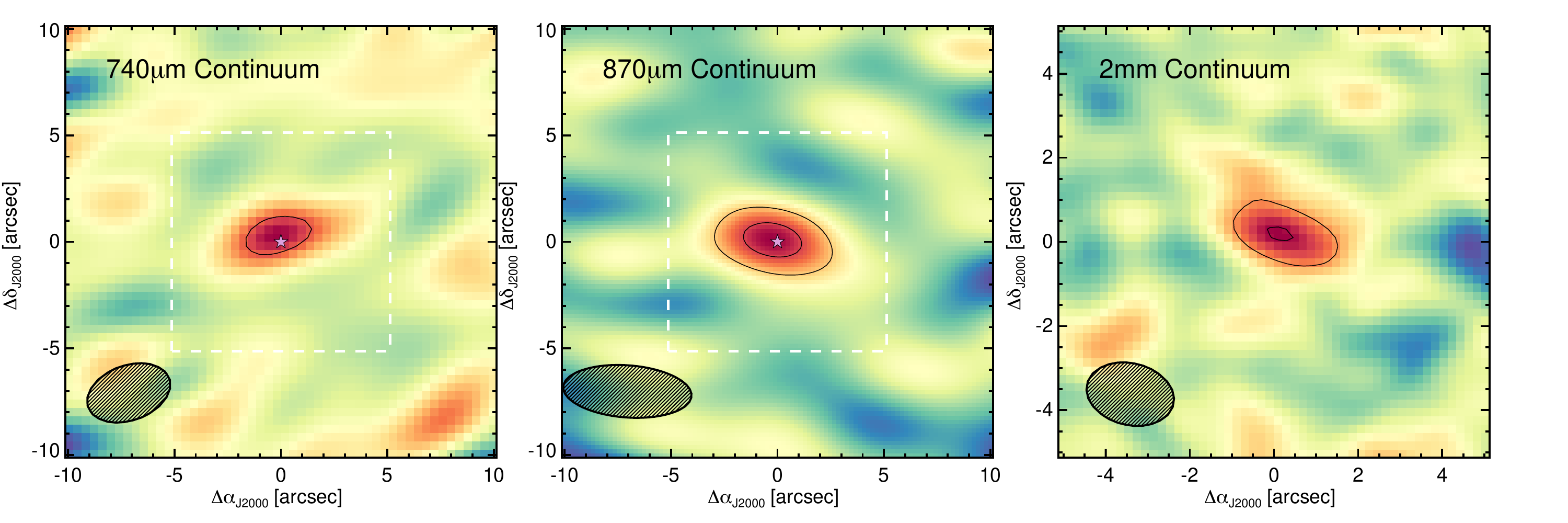}
    \caption{$20\arcsec\times20\arcsec$ and $10\arcsec\times10\arcsec$ cutouts of 740\,\micron\ (ACA), 870\,\micron\ (ACA) and 2\,mm continuum imaging for MORA-9. Contours represent 3, 5\,$\sigma$ in all three images. The star denotes the 2\,mm ALMA position of MORA-9 and the white dashed box depicts the $10\arcsec\times10\arcsec$ cutout to emphasize the varying scales between the  images. Briggs weighting (robust$=0$) is used to create the 870\,\micron\ image to maximize the spatial resolution of the ACA data while natural weighting (robust$=2$) is used for the 740\,\micron\ and 2\,mm images with lower signal-to-noise.}
    \label{fig:source9_maps}
\end{figure*}

\subsection{ALMA Data}
The 2\,mm data were obtained during Cycle 6 under program 2018.1.00231.S (PI Casey). MORA-5 and MORA-9 both sit in the contiguous 156\,arcmin$^2$ mosaic, where the root mean square (RMS) reached 89\,\uJy\ and 74\,\uJy\ at their respective source positions. We select all pointings contributing more than 5\% to the total sensitivity at the given positions (21 pointings total for both sources) before reducing and imaging the data using the Common Astronomy Software Application\footnote{http://casa.nrao.edu} (CASA) version 5.6.1 following the standard reduction pipeline scripts. We experimented with different manually defined clean boxes during the cleaning process, of order the size of the emission. We adopt a 1\arcsec\ aperture centered on the source position as our clean box, encompassing the peak of emission without including any noise. This is a slightly different reduction independent from the large mosaic presented in \cite{Zavala2021ApJ...909..165Z}, customized for the individual sources given our ability to test multiple setups, but the results are consistent with the full mosaic. 

Band 4 (2\,mm) observations covered frequencies 139.4--143.2\,GHz and 151.5--155.2\,GHz with a restoring beam of $1\farcs67\times1\farcs41$. The continuum RMS over the 7.5\,GHz bandwidth reached 85.1\,\uJy/beam. Various visibility weights were explored for imaging as we experimented with both natural and Briggs weighting (robust=0.0, 0.5, 2.0), the latter in an attempt to detect any spatially resolved components, though in the end we determined both sources to be unresolved. No spectral lines were detected in the covered bandwidth after close inspection. 
Based on the continuum imaging, the aggregate band 4 (central frequency of 147.3\,GHz) flux densities of MORA-5 and MORA-9 are $610\pm90\,$\uJy\ and $380\pm70$\,\uJy, respectively.

Band 6 (1.3\,mm) and band 7 (870\,\micron) continuum data exists from other ALMA programs 2016.1.00279.S (PI Oteo) and 2016.1.00463.S (PI Matsuda) for MORA-5. These data were downloaded from the archive and re-imaged. Band 7 observations covered frequencies 335.6--339.6\,GHz and 347.6--351.4\,GHz with a restoring beam of 0.94\arcsec$\times$0.84\arcsec. Band 6 observations covered frequencies 223--227\,GHz and 239--243\,GHz with a restoring beam of 0.90\arcsec$\times$0.76\arcsec. The emission in both bands is unresolved and the reported band 6 and band 7 flux densities for MORA-5 are $2.26\pm0.17$\,mJy and $6.49\pm0.22$\,mJy.

Comparatively, MORA-9 lacks coverage at other frequencies, so we secured follow-up Atacama Compact Array (ACA) observations under program 2019.2.00143 (PI Casey) to obtain band 6, band 7, and band 8 observations for improved dust continuum constraints. 
The band 6 observations resulted in a detection just under $5\,\sigma$ with flux density of $1.24\pm0.25$\,mJy. The aggregate band 7 continuum has a flux density of 2.54$\pm$0.37\,mJy. This is consistent with SCUBA-2 data (see Section \ref{subsec:ancillary}), but more precise both astrometrically and in the measurement of its flux density. 
The band 8 continuum flux density for MORA-9 is $3.62\pm0.66$\,mJy. Given the lower resolution of the ACA data (${\rm FWHM}=5\farcs5\times3\farcs3$) we adopt a 2\arcsec\ aperture centered on the source position as our clean box. 
Figures \ref{fig:source5_maps} and \ref{fig:source9_maps} show the ALMA continuum maps for MORA-5 and MORA-9. Natural weighting (robust$=2$) is used for all MORA-5 imaging as well as the 740\,\micron\ (because of its higher RMS) and 2\,mm continuum map of MORA-9. Briggs weighting (robust$=0$) is utilized for the 870\micron\ map of MORA-9 to maximize spatial resolution since the ACA data mitigates any loss in flux expected by not using natural weighting. 

\begin{figure*}
    \centering
    \includegraphics[width=\textwidth]{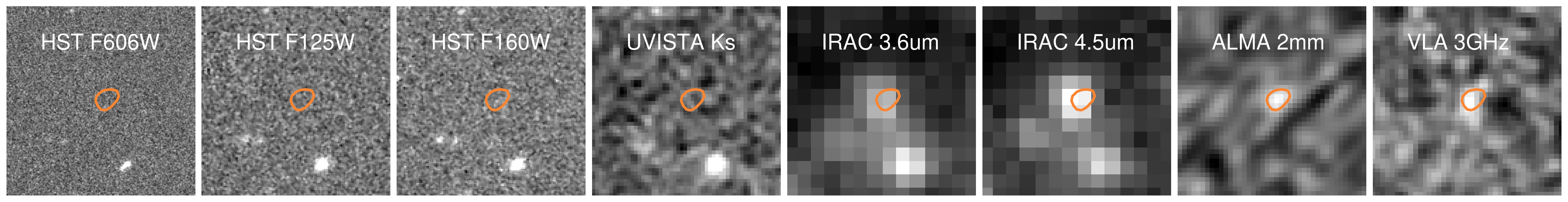}
    \includegraphics[width=\textwidth]{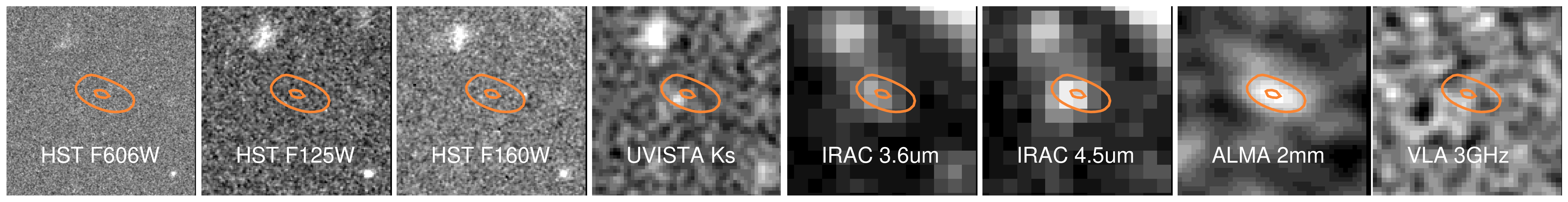}
    \caption{$8\arcsec\times8$\arcsec\ multiwavelength cutouts of MORA-5 (top) and MORA-9 (bottom) from {\it HST} ACS/F606W, {\it HST} WFC3/F125W, {\it HST} WFC3/F160W, Ultra-VISTA {\it Ks}-band, {\it Spitzer} 3.6\,\micron\ and 4.5\,\micron, ALMA 2\,mm, and VLA 3\,GHz. Orange contours in each frame show the 5\,$\sigma$ contours (3 and 5\,$\sigma$ for MORA-9) from the 2\,mm imaging.}
    \label{fig:cutouts}
\end{figure*}

\subsection{Ancillary Archival COSMOS Datasets}
\label{subsec:ancillary}
Both sources lie in the Cosmic Evolution Survey Field \citep[COSMOS;][]{Scoville2007ApJS..172....1S,Capak2007ApJS..172...99C} within the Cosmic Assembly Near-infrared Deep Extragalactic Legacy Survey \citep[CANDELS;][]{Koekemoer2011ApJS..197...36K,Grogin2011ApJS..197...35G} area. 
We explore a range of apertures centered on the ALMA 2\,mm positions and extract photometry from the available OIR data. Apertures of 1.5\arcsec\ diameter are placed down on the {\it HST} F606W, F814W, F125W, F140W, and F160W images resulting in no detections. 

While neither source was initially found to have counterparts in the original $Ks$-selected catalog \citep{Muzzin2013aApJS2068M} within 1\arcsec, a $Ks$-band detection for MORA-9 is now reported in the updated COSMOS2020 catalog (Weaver et al. submitted) after the inclusion of the fourth data release (DR4) of the UltraVISTA survey \citep{McCracken2012} which folds in deeper (up to 1 mag) $Ks$-band imaging. MORA-5 however, still lacks a robust detection down to the 25.2\,mag ($3\,\sigma$) limit with a signal-to-noise ratio (SNR) $\sim1.5$. 

Both sources do have detections in the deep S-CANDELS {\it Spitzer} IRAC data \citep{Ashby2015ApJS21833A} at 3.6\,\micron\ and 4.5\,\micron. Neither MORA-5 nor MORA-9 are detected in {\it Spitzer} 24\,\micron\ imaging \citep{LeFloch2009}, {\it Herschel} PACS 100\,\micron\ and 160\,\micron\ imaging \citep{Lutz2011}, or SPIRE 250\,\micron\,, 350\,\micron\, and 500\,\micron\ imaging \citep{Oliver2012}.

The sources were also both detected with SCUBA-2 as reported in \cite{Geach2017} and then again in the deeper, expanded S2COSMOS survey presented in \cite{Simpson2019ApJ88043S}. The ratio of 850\,\micron\ flux density to 2\,mm flux density for both MORA-5 and MORA-9 is what initially indicated these sources were likely to be high redshift at $z\geq3$ ($S_{850\mu{\rm m}}/S_{2{\rm mm}}=11.1\pm$2.5 and 6.9$\pm$2.8, respectively). Specifically, the ratio for MORA-9 is significantly lower than the expected value for emission on the Rayleigh-Jeans tail of the cold dust blackbody. This suggests that the SCUBA-2 measurement at 850\,\micron\ is near the peak of the dust SED rather than lower down on the Rayleigh-Jeans tail.

Lastly, neither source was found to have a detection significant enough to be included in the 1.4\,GHz VLA-COSMOS catalog ($3\,\sigma\sim30\,$\uJy\,beam$^{-1}$; \citealt{Schinnerer2007ApJS..172...46S}) or the 3\,GHz VLA-COSMOS catalog ($5\,\sigma\sim12\,$\uJy\,beam$^{-1}$; \citealt{Smolcic2017}). After manually extracting fluxes from the radio imaging at the 2\,mm source positions, we find marginal detections of $S_{3{\rm GHz}}=10.1\pm2.4\,$\uJy\ for MORA-5 and $S_{3{\rm GHz}}=4.3\pm3.4\,$\uJy\ for MORA-9. 

\begin{table*}
\centering
\caption{Photometry of MORA-5 and MORA-9}
\label{tab:photmetry}
\begin{tabular}{cccccc}
\hline
\hline
Band & Wavelength & Units & MORA-5 & MORA-9 & Reference \\
\hline
{\it HST}-F606W & 606\,nm & nJy & ($-11\pm$24) & (50$\pm$24) & \cite{Koekemoer2011ApJS..197...36K} \\
{\it HST}-F814W & 814\,nm & nJy & (8$\pm$32) & ($-21\pm$32) & \cite{Koekemoer2011ApJS..197...36K}\\
{\it HST}-F125W & 1.25\,\micron & nJy & ($-12\pm$44) & ($-28\pm$43) & \cite{Koekemoer2011ApJS..197...36K}\\
{\it HST}-F140W & 1.40\,\micron & nJy & (145$\pm$83) & -- & \cite{Brammer2012ApJS..200...13B}\\ 
{\it HST}-F160W & 1.60\,\micron & nJy & (40$\pm$42) & (40$\pm$43) & \cite{Koekemoer2011ApJS..197...36K}\\
UVISTA-Ks & 2.2\,\micron & nJy & (59$\pm$40) & 269$\pm$40 & \citealt{McCracken2012} \\
IRAC-CH1 & 3.6\,\micron & nJy & 680$\pm$140 & 760$\pm$150 & \cite{Ashby2015ApJS21833A}\\
IRAC-CH2 & 4.5\,\micron & nJy & 790$\pm$160 & 880$\pm$170 & \cite{Ashby2015ApJS21833A} \\
MIPS24 & 24\,\micron & \uJy & (1$\pm$20) & ($-2\pm$10) & \cite{LeFloch2009} \\
PACS & 100\,\micron & mJy & ($-0.2\pm$1.7) & ($-0.5\pm$1.7) & \cite{Lutz2011} \\
PACS & 160\,\micron & mJy & ($-0.1\pm$3.7) & ($-0.3\pm$3.7) & \cite{Lutz2011} \\
SPIRE & 250\,\micron & mJy & (3.8$\pm$5.8) & (0$\pm$5.8) & \cite{Oliver2012} \\
SPIRE & 350\,\micron & mJy & (7.3$\pm$6.3) & (0$\pm$6.3) & \cite{Oliver2012} \\
SCUBA-2 & 450\,\micron & mJy & (8.53$\pm$4.13) & -- & \cite{Casey2013} \\
SPIRE & 500\,\micron & mJy & (7.1$\pm$6.8) & (0$\pm$6.8) & \cite{Oliver2012} \\
ALMA-B8 & 740\,\micron & mJy & -- & 3.62$\pm$0.66 & THIS WORK \\
SCUBA-2 & 850\,\micron & mJy & 6.80$\pm$0.53 & 2.61$\pm$0.59 & \cite{Simpson2019ApJ88043S} \\
ALMA-B7 & 868\,\micron & mJy & 6.49$\pm$0.22 & -- & THIS WORK \\
ALMA-B7 & 874\,\micron & mJy & -- & 2.54$\pm$0.37 & THIS WORK \\
AzTEC & 1100\,\micron & mJy & 3.7$\pm$0.9 & -- & \cite{Aretxaga2011} \\
ALMA-B6 & 1287\,\micron & mJy & 2.26$\pm$0.17 & $1.24\pm0.25$ & THIS WORK \\
ALMA-B4 & 2036\,\micron & \uJy & 610$\pm$90 & 380$\pm$70 & THIS WORK \\
VLA-3\,GHz & 10\,cm & \uJy & (10.1$\pm$3.4) & (4.3$\pm$2.4) & \cite{Smolcic2017} \\
VLA-1.4\,GHz & 21.4\,cm & \uJy & ($20.0\pm12.5$) & ($-4.2\pm10.8$) & \cite{Schinnerer2007ApJS..172...46S} \\
\hline
\hline
\end{tabular}
\begin{flushleft}
{\bf Notes --} Available photometric measurements for MORA-5 and MORA-9. Non-detections and measurements with $<3\sigma$ significance are denoted by parenthesis. MORA-9 sits just outside the coverage area of the {\it HST}-F140W observations. Neither source is detected in the intermediate band filters available in COSMOS. ALMA band 7 and band 6 were downloaded from the archive and re-imaged for MORA-5. Band 6, band 7, and band 8 data for MORA-9 were obtained with follow-up ACA observations. 
\end{flushleft}
\end{table*}

\section{SED Fitting Techniques}
\label{sec:sed_fitting_techniques}
We explore the use of several spectral energy distribution (SED) fitting techniques in relation to the available multiwavelength data on MORA-5 and MORA-9 to derive redshift constraints and physical properties. We elect to input the photometry exactly as it appears in Table \ref{tab:photmetry} into our SED fitting, and therefore our photometric redshift estimates, preserving all formal non-detections rather than substituting them with upper limits. We also note here for clarity that the final redshift estimates adopted for MORA-5 and MORA-9 are a combination of the photometric redshift probability distribution functions (PDFs) of each technique and are referred to as $z_{\rm PDF}$ throughout the paper. Due to the highly obscured nature of these galaxies, it would be possible to have an extremely obscured, optically thick central region. Implementing energy balance techniques in SED fitting works best when data sampling the full SED, both the OIR and long wavelength regimes, is available and the galaxy in question does not have much optically thick dust. For these reasons we fit the SEDs of the OIR and FIR/mm regimes separately, but also fit the full SED for a fair comparison of the differing techniques. We recognize the challenge of fitting SEDs without well constrained redshifts, and as we are lacking spectroscopic redshifts for MORA-5 and MORA-9, we approach the process iteratively. Here we describe the tools used for photometric redshift and SED fitting and Section \ref{subsec:photo-zs} describes our results with respect to redshift constraints.

\begin{figure*}
\centering
\begin{minipage}{0.48\textwidth}
\includegraphics[width=0.99\textwidth]{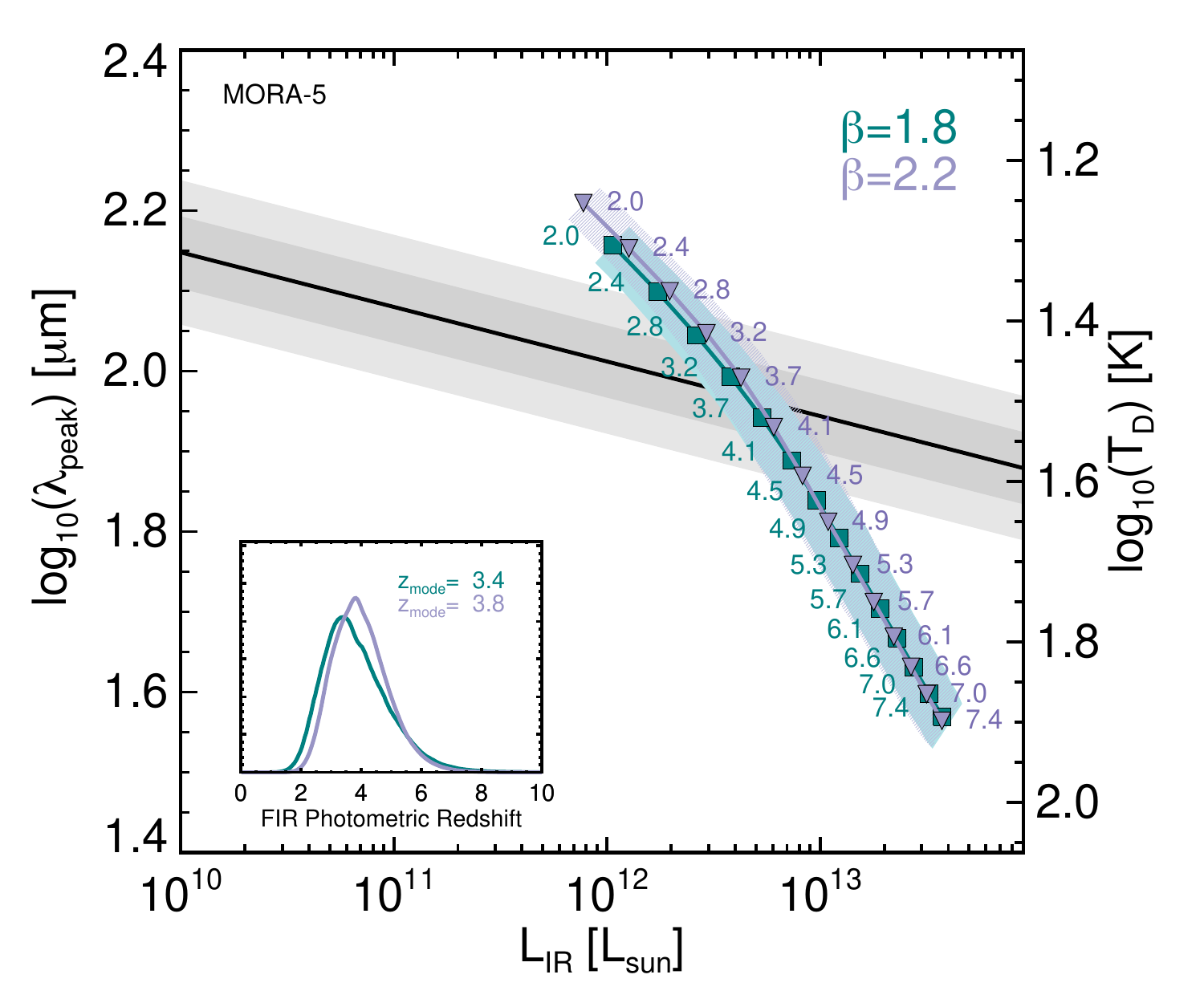}
    \end{minipage}
\begin{minipage}{0.48\textwidth}
\includegraphics[width=0.99\textwidth]{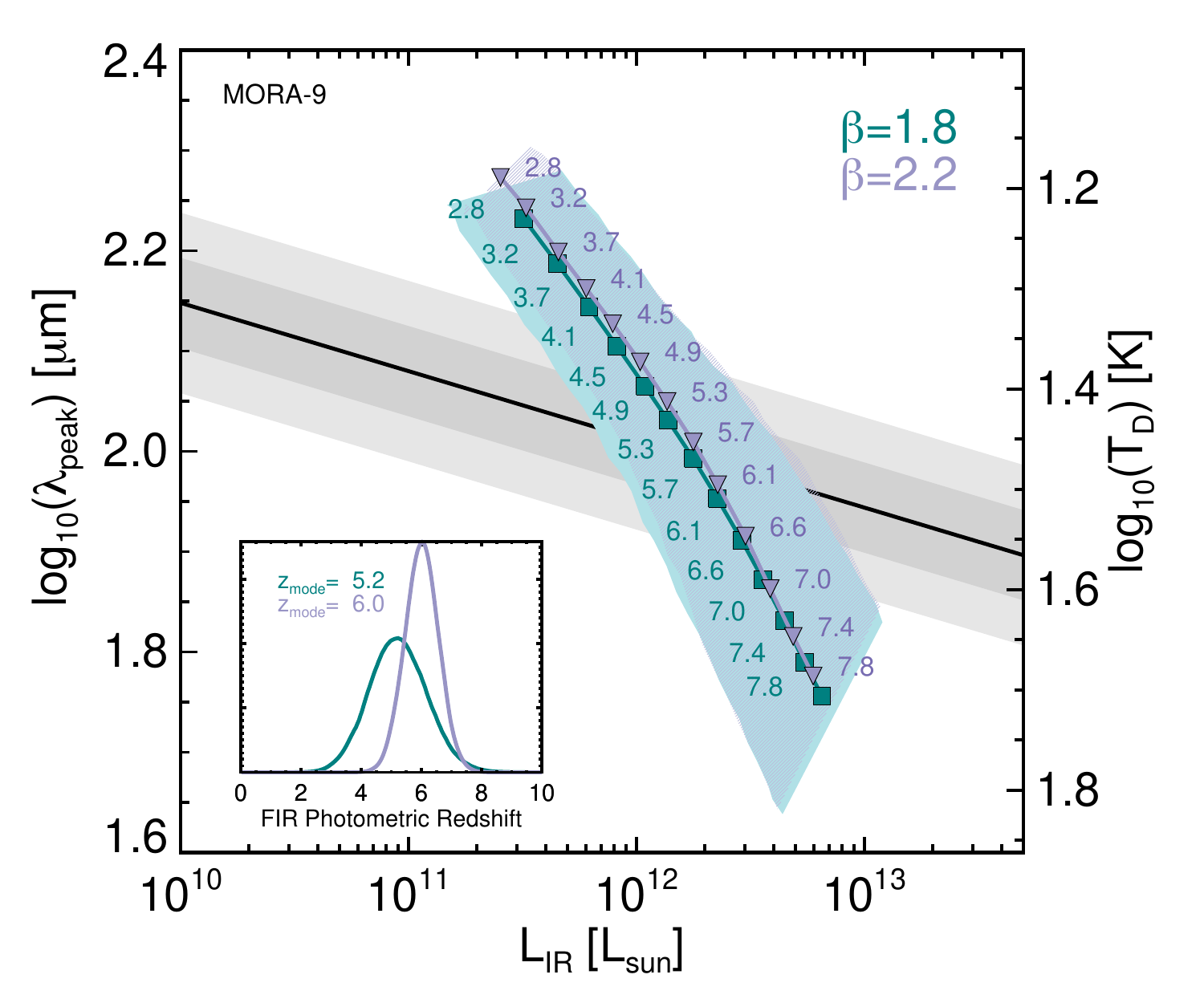}
\end{minipage}
\caption{L$_{\rm IR}$--$\lambda_{\rm peak}$ tracks traced by SEDs fit to the available photometry with associated uncertainty from $z=2-8$. We show results using $\beta=1.8$ (teal) and 2.2 (purple) to assess if the value of $\beta$ affects where these tracks fall in the L$_{\rm IR}$--$\lambda_{\rm peak}$ parameter space. Inset plots show the corresponding photometric redshift distributions from the results of {\sc MMpz}. The {\sc MMpz} redshift is dictated by the intersection of the SED fit tracks and the intrinsic L$_{\rm IR}$--$\lambda_{\rm peak}$ relation in galaxies (black line with 1,2\,$\sigma$ scatter in grey).}
\label{fig:lpeak_vs_lir_mmpz}
\end{figure*}

\subsection{FIR/millimeter SED and {\sc MMpz} }
\label{subsec:mmpz}
We fit the FIR/mm obscured SED with a piecewise function comprised of a single modified blackbody and mid-IR powerlaw using a Metropolis Hastings Markov Chain Monte Carlo (MCMC). For a complete description of the FIR/mm SED fitting technique, called MCIRSED, we encourage the reader to see the forthcoming publication by P. Drew et al. in prep., but briefly explain the model here. At short wavelengths, the FIR/mm SED is dominated by a power law with mid-IR slope, $\alpha$, representing warmer dust emission from star-forming regions and/or active galactic nuclei (AGN). At longer wavelengths, a modified Planck function dominates the SED and represents the cold dust. An MCMC is then employed to sample the posteriors of the fit parameters and provide confidence intervals. 
We fix the mid-IR power law slope, $\alpha=4$, and emissivity spectral index, $\beta=1.8$, following \cite{Casey2019ApJ88755C}. We also fix the wavelength where opacity equals unity, $\lambda_0$, to $200$\,\micron\ as is commonly adopted in the literature. Finally, we leave the dust temperature ($T_{\rm dust}$), wavelength corresponding to the peak in the FIR/mm SED ($\lambda_{\rm peak}$), and IR luminosity ($L_{\rm IR(8-1000\mu m)}$) as free parameters. 

\cite{2018ApJ...862...77C} (see Figure 3 therein) finds an empirical relationship between galaxies' intrinsic $L_{\rm IR(8-1000\mu m)}$ and their observed $\lambda_{\rm peak}$ given the typical SEDs of cold dust emission in DSFGs. Comparing these two properties provides us with an approximation for the range of redshift solutions which are feasible based on existing observations. This empirical relationship is the basis for the {\sc MMpz} fitting technique \citep{Casey2020ApJ...900...68C} which we use to obtain FIR photometric redshift estimates. The results of this method are illustrated in Figure \ref{fig:lpeak_vs_lir_mmpz}. The  $L_{\rm IR}$--$\lambda_{\rm peak}$ trend is shown by the black line with its associated 1 and 2\,$\sigma$ scatter from measured data in grey. Across a broad span of redshifts, the teal and purple tracks trace the range of possible SEDs constrained by the measured photometry given $\beta$ values of 1.8 and 2.2 (fixed only to $\beta=1.8$ for simplicity in \citealt{Casey2020ApJ...900...68C}). As we can see, the choice of $\beta$ has a negligible effect on the redshift solution, so we adopt the $\beta=1.8$ value for consistency. The associated errors reflect the number of photometric data points available to sample the SEDs, resulting in larger errors for MORA-9. 

\subsection{{\sc eazy} OIR SED}
\begin{figure*}
    \centering
    \includegraphics[width=0.95\textwidth]{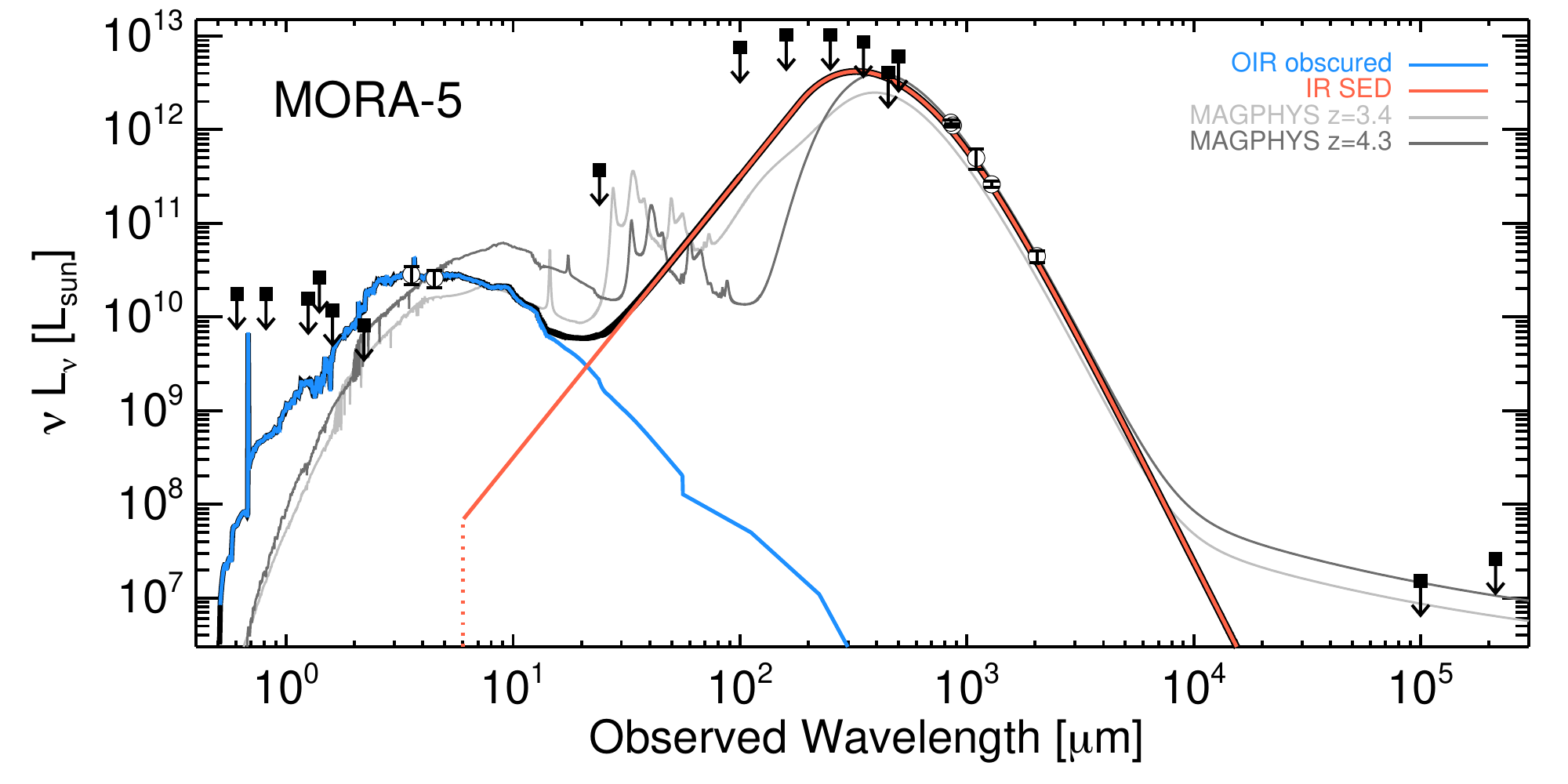}
    \includegraphics[width=0.95\textwidth]{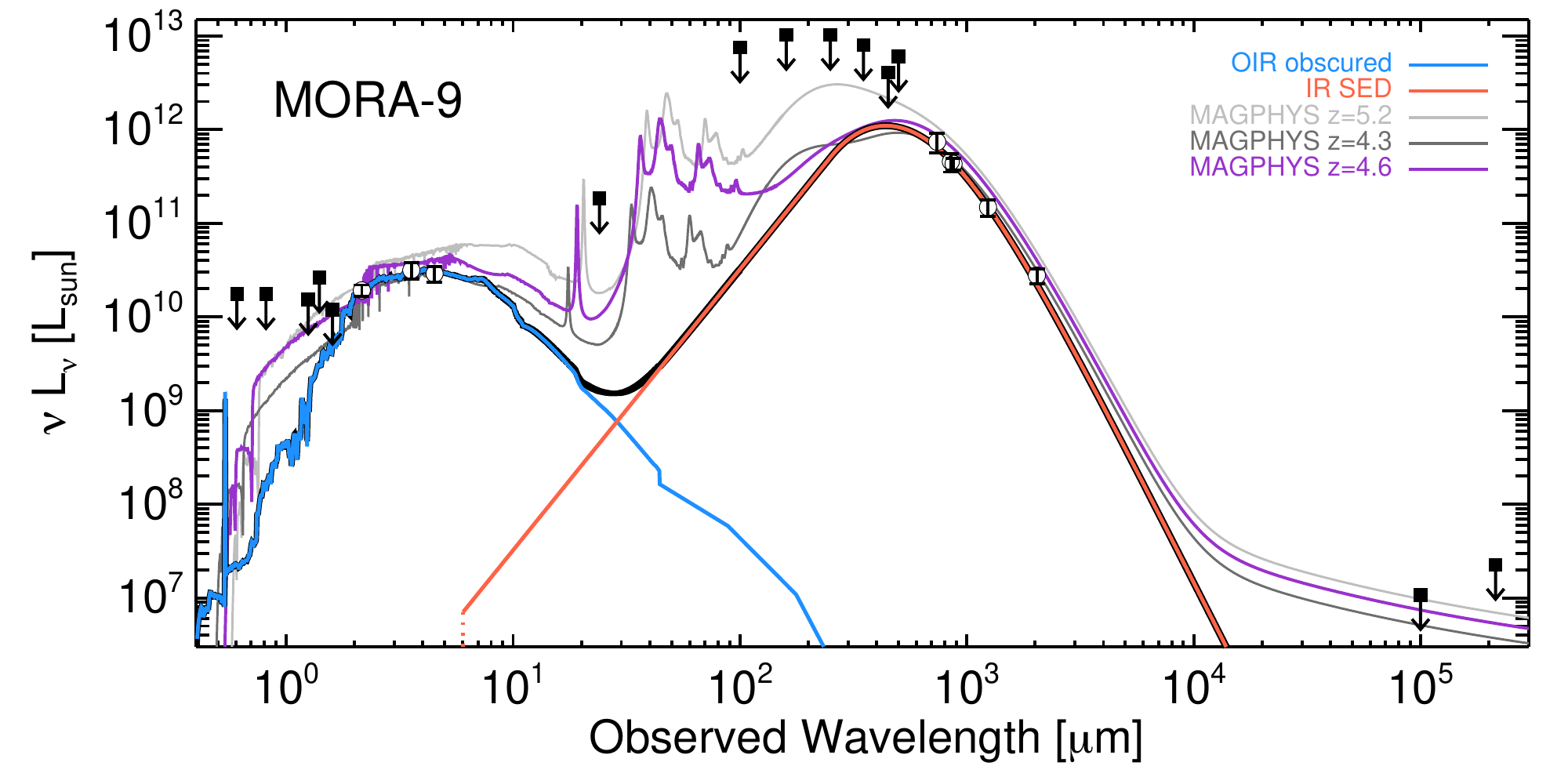}
    \caption{Composite SED (black) comprised of two components: OIR stellar emission (blue) from {\sc eazy} and thermal dust emission from {\sc MMpz} (orange). Black arrows depict $3\,\sigma$ upper limits for illustrative purposes, while the exact photometric values listed in Table \ref{tab:photmetry} are used for SED creation. Light and dark grey lines are {\tt MAGPHYS} SEDs, having provided the original {\tt MAGPHYS} with {\sc MMpz} $z_{\rm mode}$ (light grey) and $z_{\rm PDF}$ (dark grey) as input redshifts. Note that the offset between {\tt MAGPHYS} SEDs is due to the conversion from flux density to $\nu\,L_{\nu}$ being dependent on luminosity distance (i.e. redshift). The OIR SED, FIR/mm SED, data points, and upper limits have been converted given the luminosity distance at \zpdf. {\it Bottom:} The purple line shows the {\tt MAGPHYS} SED of MORA-9 based on the redshift reported in the COSMOS2020 catalog.}
    \label{fig:full_seds}
\end{figure*}

\begin{figure*}
    \centering
\begin{minipage}{0.49\textwidth}
\includegraphics[width=\textwidth]{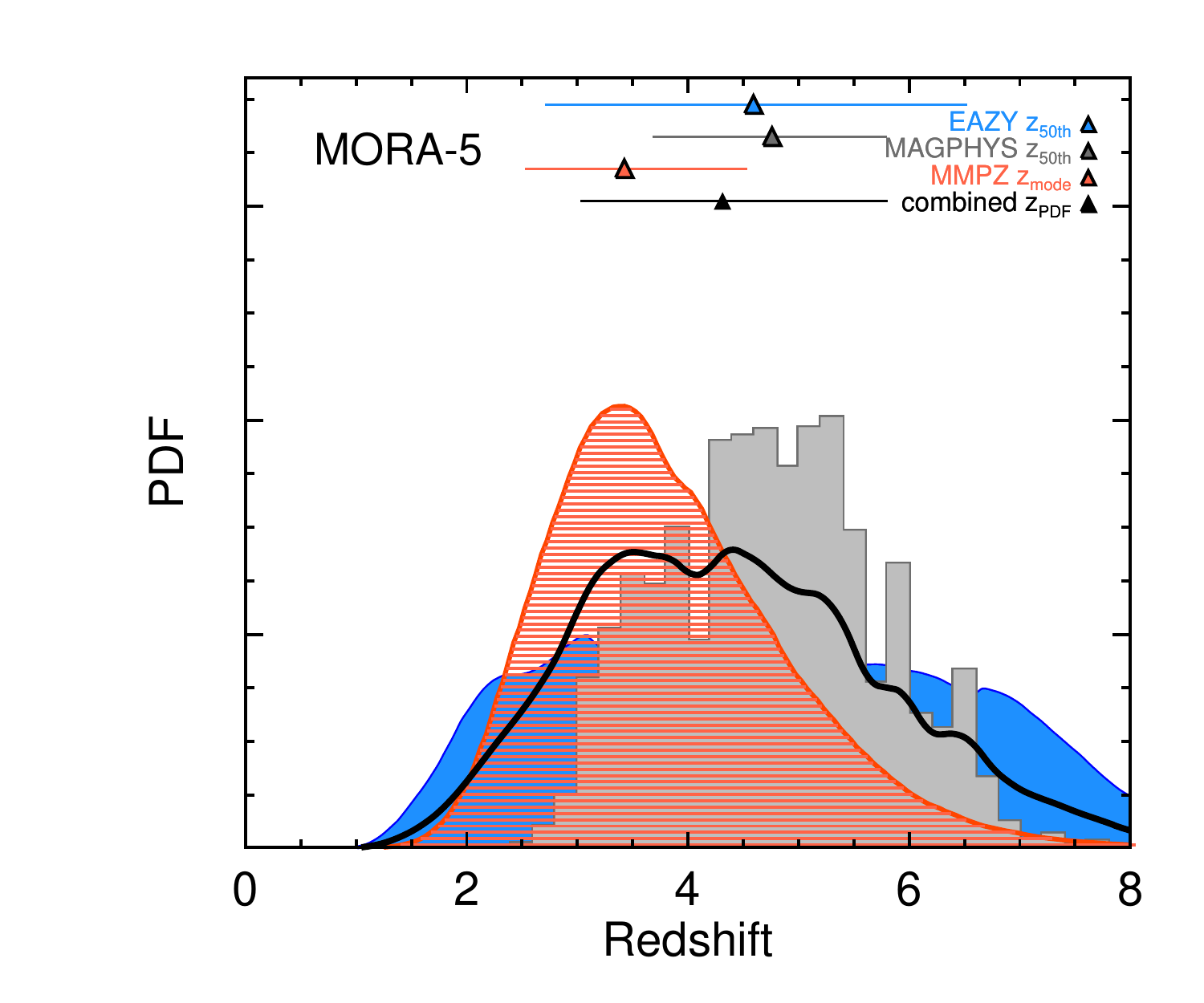}
    \end{minipage}
\begin{minipage}{0.49\textwidth}
\includegraphics[width=\textwidth]{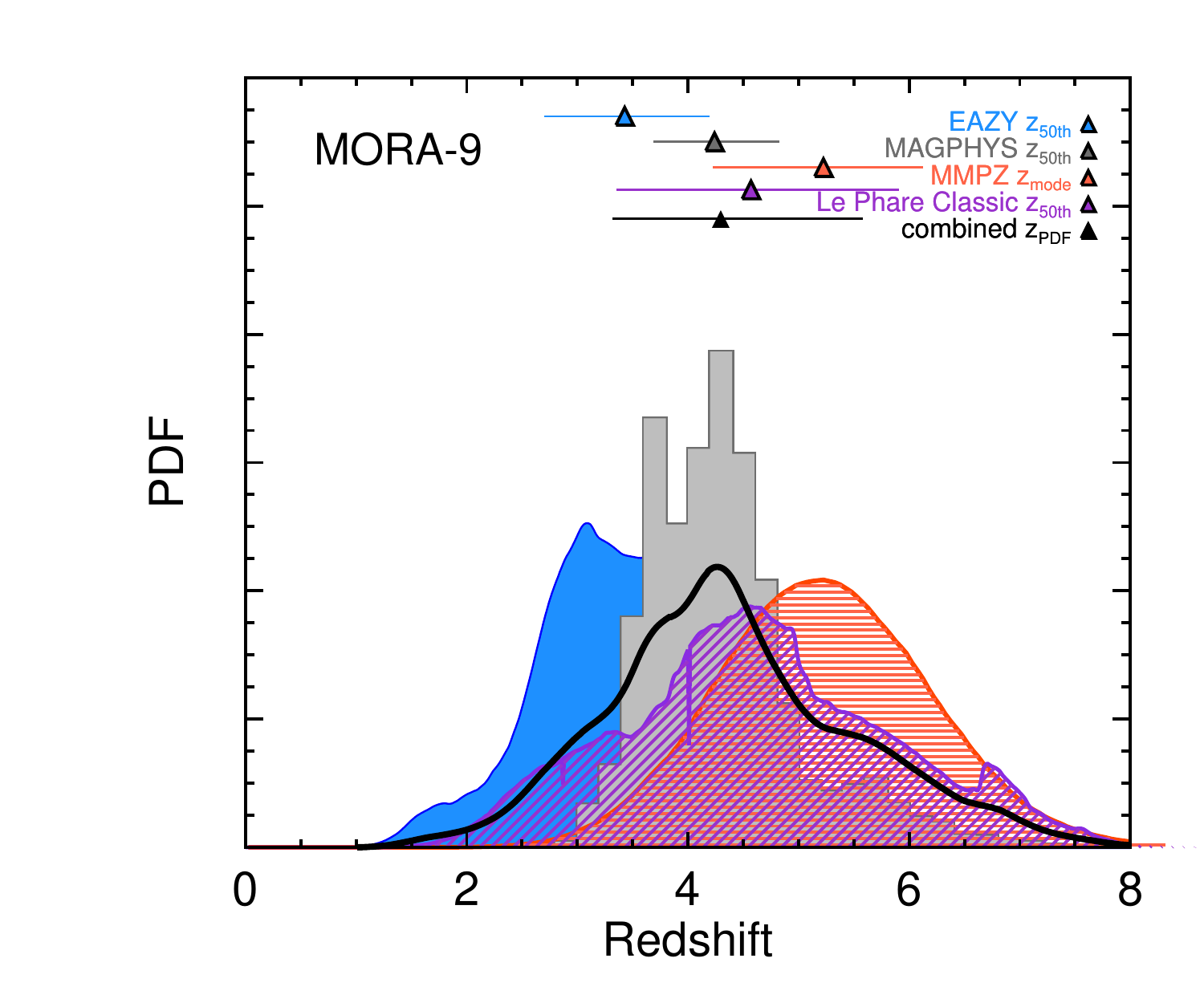}
    \end{minipage}
\caption{Probability distribution functions (PDFs) of the three methods used to determine photometric redshifts for MORA-5 (left) and MORA-9 (right): OIR photometry with {\sc eazy} (blue), energy balance technique with \magphysphotz\ (gray), and FIR/mm photometry with {\sc MMpz} (orange). For MORA-9 (right), we also show the PDF result from the LePhare template fitting code \citep{Ilbert2006A&A...457..841I} following the same method as \cite{Laigle2016ApJS..224...24L} as reported in the COSMOS2020 catalog (magenta). $1\,\sigma$ errors are shown at the top with triangles denoting the photometric redshifts. 
For our {\sc eazy} and \magphysphotz\ (shortened in the legend to {\sc magphys} for space) results, $z_{\rm 50th}$ signifies the redshifts determined from marginalizing over their respective distributions. We report the final redshift estimates ($z_{\rm PDF}$) for these sources as the 50th percentile value of the combined distribution of all three (four for MORA-9) PDFs, illustrated by the black lines.}
\label{fig:zphot_pdfs}
\end{figure*}

We fit the sparse OIR data with the photometric redshift fitting software \textsc{eazy} (v1.3) \citep{Brammer2008ApJ686} using the provided set of templates. The eight \textsc{eazy} templates result from a linear combination of $\sim500$ \cite{Bruzual2003MNRAS.344.1000B} synthetic galaxy photometry models and includes a dusty starburst model, an important addition for this work. Due to the limited OIR data, particularly for MORA-5, our results show very broad photometric redshift probability distributions as expected. The \textsc{eazy} SEDs are shown as blue lines in Figure \ref{fig:full_seds} and their associated PDFs correspond to the blue distributions in Figure \ref{fig:zphot_pdfs}. 

\subsection{{\tt MAGPHYS} Energy Balance SED}
Lastly, the full SED is fit using {\tt MAGPHYS} and the updated \magphysphotz\ code \citep{DaCunha2008MNRAS3881595D,DaCunha2015ApJ806110D,Battisti2019ApJ...882...61B}. The original {\tt MAGPHYS} incorporates an energy balance technique taking into account emission from stellar populations as well as absorption and emission by dust in galaxies. We opt to use \magphysphotz\ results rather than the classic {\tt MAGPHYS} 
given its inclusion of IR, (sub)mm, and radio data, plus new star-formation history (SFH) and dust temperature priors which are more appropriate for high redshift DSFGs. Specifically, we adopt stellar masses and rest-frame $V$-band dust attenuation (A$_V$) values from this updated version. However, we obtain the SEDs shown in Figure \ref{fig:full_seds} by running the original {\tt MAGPHYS}, setting the redshift to that determined from {\sc MMpz} as well as the final $z_{\rm PDF}$. Light gray lines are the {\tt MAGPHYS} SEDs given $z_{\rm mode}$ from {\sc MMpz} while dark gray shows the {\tt MAGPHYS} SED assuming the $z_{\rm PDF}$. For MORA-9, we also run {\tt MAGPHYS} given the COSMOS2020 redshift and this is shown as the purple line in the bottom panel. We note that the SEDs produced by \magphysphotz\ are also consistent with the original {\tt MAGPHYS} SEDs when \zpdf\ is provided as the redshift. Thus, we do not show the \magphysphotz\ SEDs in Figure \ref{fig:full_seds}. 

\section{Results}
\label{sec:results}
We rely on photometry to estimate redshifts and several physical properties for MORA-5 and MORA-9. Here we discuss these physical properties, based on the \zpdf\ derived from the various SED fitting techniques and computed in Section \ref{subsec:photo-zs}, and list them in Table \ref{tab:derived_properties}.

\subsection{Photometric Redshifts}
\label{subsec:photo-zs}
Photometric redshift PDFs from all SED fitting techniques utilized are displayed in Figure \ref{fig:zphot_pdfs}. The redshift determined from each one and associated $1\,\sigma$ errors are depicted by the triangles at the top of the plot. We sum all available PDFs (three for MORA-5 and four for MORA-9 after the inclusion of the COSMOS2020 distribution) and adopt the value from marginalizing over the PDF as our final photometric redshift which we refer to as $z_{\rm PDF}$ (black distribution). Combining photometric redshift PDFs from different codes in this way has proven to be advantageous \citep{Dahlen2013ApJ...775...93D}. For MORA-5, $z_{\rm PDF}=4.3^{+1.5}_{-1.3}$ and for MORA-9, $z_{\rm PDF}=4.3^{+1.3}_{-1.0}$. The available ALMA observations only offer a few narrow frequency ranges for molecular lines to be detected and a detection at these frequencies would be indicative of a $z<4$ redshift solution, providing some evidence for our adopted estimates. Next, we describe the PDFs and associated photometric redshifts which contribute to the \zpdf\ values. 

From {\sc MMpz} (orange distribution), $z_{\rm mode}$ is the mode of the PDF derived from the intersection of the galaxy's redshift track with the $L_{\rm IR}$--$\lambda_{\rm peak}$ trend from the literature as shown in Figure \ref{fig:lpeak_vs_lir_mmpz}. For MORA-5, $z_{\rm mode}=3.4^{+1.1}_{-0.9}$ and for MORA-9, $z_{\rm mode}=5.2^{+0.9}_{-1.0}$. This is the highest redshift estimate for MORA-9. Low-redshift solutions for this source would imply implausibly cold dust for $z\lesssim3$ ($\lambda_{\rm peak}>163$\,\micron\ or T${\rm dust}<18$\,K). This would be an extremely cold system considering the luminosity of MORA-9 ($L_{\rm IR}\sim10^{12}\,L_\odot$, which is largely insensitive to redshift) and as such we find a higher redshift solution to be more likely given its FIR-mm colors.

Results from {\sc eazy} are illustrated in blue in both panels of Figure \ref{fig:zphot_pdfs}. We show the redshifts 
determined from marginalizing over the full {\sc eazy} PDF ($z_{\rm 50th}=4.6\pm1.9$ for MORA-5 and $z_{\rm 50th}=3.4^{+0.8}_{-0.7}$ for MORA-9). The broad distribution of the \textsc{eazy} PDF is expected given the limited OIR photometric data, particularly for MORA-5 which lacks a $Ks$-band detection and is only constrained by two IRAC detections.

The photometric redshift distributions from \magphysphotz\ are displayed in grey. Similar to our {\sc eazy} results, we show the 50th percentile redshift values for each source. For MORA-5 the redshift estimate is $z_{\rm 50th}=4.8^{+1.0}_{-1.1}$, while for MORA-9 it is $z_{\rm 50th}=4.2\pm0.6$.

As noted in Section \ref{sec:data_and_obs}, MORA-9 was found to have a $Ks$-band counterpart in the most recent UltraVISTA data and consequently is included in the COSMOS2020 catalog with a corresponding redshift. COSMOS2020 includes photometric redshift estimates using both {\sc eazy} and the same method as cited in \cite{Laigle2016ApJS..224...24L} which utilizes the template-fitting code {\tt LePhare} \citep{Arnouts2002MNRAS.329..355A,Ilbert2006A&A...457..841I}. We refer the reader to Weaver, Kauffmann et al. (submitted) for a full explanation, but include these results for comparison and fold in the COSMOS2020 PDF into the final $z_{\rm PDF}$ estimate for MORA-9. The redshift reported for MORA-9 in the Classic {\tt LePhare} COSMOS2020 catalog is $z_{\rm 50th}=4.57^{+0.87}_{-0.89}$.

\subsection{IR Luminosity and Star Formation Rate}
Total infrared luminosities ($L_{\rm IR}$) are determined by integrating the {\sc MMpz} FIR SEDs from 8--1000\,\micron. In Table \ref{tab:derived_properties} we list the $L_{\rm IR}$ assuming the combined \zpdf\ values. With these values we then calculate star-formation rates (SFRs) utilizing calibrators from the literature (\citealt{Kennicutt&Evans2012ARA&A50531K} with references to \citealt{Murphy2011ApJ73767M,Hao2011ApJ741124H} therein) and obtain SFR estimates of 830$^{+34}_{-190}$\,M$_\odot$\,yr$^{-1}$ for MORA-5 and 200$^{+250}_{-60}$\,M$_\odot$\,yr$^{-1}$ for MORA-9. If we were to assume the {\sc MMpz}-determined redshifts and corresponding $L_{\rm IR}$ instead, the SFRs shift to 680$^{+140}_{-120}$\,M$_\odot$\,yr$^{-1}$ and 340$^{+350}_{-100}$\,M$_\odot$\,yr$^{-1}$ respectively, remaining consistent to the previous SFRs given the large uncertainties. Since the IR luminosities are redshift dependent, the SFRs increase if we assume higher redshift solutions. For MORA-5, {\sc MMpz} produces a lower redshift than the result of the combined PDFs, whereas MORA-9 is the inverse case and {\sc MMpz} produces a higher redshift. Had we used the least constrained photometric redshifts produced by {\sc eazy}, the opposite would occur: The higher redshift estimate of MORA-5 would push it to an extreme IR luminosity ($\sim10^{13}$\,L$_\odot$) and subsequent SFR, while the lower redshift estimate of MORA-9 would suggest the system more closely resembles a normal, luminous infrared galaxy \citep[LIRG:][]{SandersMirabel1996ARA&A..34..749S} found in the local Universe with $L_{\rm IR}\sim5\times10^{11}$\,L$_\odot$.

\subsection{Stellar Mass and Attenuation}
We report the stellar masses ($M_\star$) and absolute magnitudes of extinction ($A_V$) values inferred from \magphysphotz. As shown in \cite{Battisti2019ApJ...882...61B}, M$_\star$ can be underestimated by a factor of two for DSFGs, and is most divergent at higher stellar masses ($\gtrsim10^{10}$\,M$_\odot$) when using codes which rely only on rest-frame UV-NIR data such as {\sc eazy} and the original {\tt MAGPHYS}. Given this, we report the stellar masses for MORA-5 and MORA-9 to be (1.5$^{+1.0}_{-0.7})\times10^{11}$\,M$_\odot$ and (4.1$^{+1.8}_{-1.4})\times10^{10}$\,M$_\odot$ while $A_V$ is 4.3$^{+0.9}_{-0.7}$ and 2.9$^{+0.4}_{-0.5}$, respectively. We note that both estimates are only marginally dependent on redshift within each galaxy's PDF. Any further improvements in our stellar mass estimates will require better constraints on both redshifts and the near/mid-IR SED from, for example, future JWST observations. 

\begin{table*}
\centering
\caption{Derived Properties of MORA-5 and MORA-9}
\label{tab:derived_properties}
\begin{tabular}{cccc}
\hline
\hline
    Property & Units & MORA-5 & MORA-9 \\
\hline
    RA & -- & 10:00:24.157 & 10:00:17.298 \\
    DEC & -- & +02:20:05.39 & +02:27:15.80 \\
    $z_{\rm PDF}$ & -- & 4.3$^{+1.5}_{-1.3}$ & 4.3$^{+1.3}_{-1.0}$ \\
    $L_{{\rm IR}(8-1000\micron)}$ & L$_{\odot}$ & (5.6$^{+2.3}_{-1.3}$)$\times10^{12}$ & (1.4$^{+1.7}_{-0.4}$)$\times10^{12}$ \\
    SFR & M$_{\odot}$yr$^{-1}$ & 830$^{+340}_{-190}$ & 200$^{+250}_{-60}$ \\
    $\lambda_{\rm peak (rest)}$ & \micron & 83$^{+9}_{-8}$ & 106$^{+19}_{-20}$ \\
    $T_{\rm dust}$ & K & 60$^{+8}_{-7}$ & 43$^{+14}_{-9}$ \\
    $M_\star$ & M$_{\odot}$ & (1.5$^{+1.0}_{-0.7})\times10^{11}$ & (4.1$^{+1.8}_{-1.4})\times10^{10}$ \\    
    $M_{\rm dust}$ & M$_{\odot}$ & (3.6$\pm$0.5)$\times10^9$ & (2.2$\pm$0.4)$\times10^9$ \\
    $M_{\rm gas}$(2\,mm) & M$_{\odot}$ & (2.6$\pm$0.4)$\times10^{11}$ & (1.6$\pm$0.3)$\times10^{11}$ \\
    $A_V$ & -- & 4.3$^{+0.9}_{-0.7}$ & 2.9$^{+0.4}_{-0.5}$ \\ 
    $q_{\rm IR}$ & -- & $>1.9$ & $>1.6$ \\
\hline
\hline
\end{tabular}
\begin{flushleft}
{\bf Notes --} Derived properties based on the separately fit OIR and FIR/mm SEDs. The \zpdf\ reported comes from combining the photometric redshift PDFs of all SED fitting techniques discussed in Section \ref{sec:sed_fitting_techniques}. 
$L_{{\rm IR}(8-1000\micron)}$ is the derived IR luminosity integrated from 8--1000\,\micron. SFRs are determined directly from our $L_{\rm IR}$ results using the associated conversion factors reported in \cite{Kennicutt&Evans2012ARA&A50531K}. $\lambda_{\rm peak}$ is the rest-frame wavelength where the FIR/mm dust SED peaks. Stellar masses and $A_V$ are taken from the \magphysphotz\ energy balance SED results. Dust and gas masses are derived directly from the 2\,mm dust continuum measurements. $q_{\rm IR}$ is computed directly from the FRC and is reported as a lower limit for MORA-5 given its marginal detection at 3\,GHz. 
\end{flushleft}
\end{table*}

\subsection{Rest-frame Peak Wavelength and Dust Temperature}
As described in Section \ref{subsec:mmpz}, $\lambda_{\rm peak}$ is the wavelength at which the best fit model peaks after fitting the FIR/mm data with a mid-IR power law plus a modified blackbody. The relationship between the measurable quantity, $\lambda_{\rm peak}$, and dust temperature, $T_{\rm dust}$, is dependent on the dust opacity model assumed (see Figure 20 of \citealt{Casey2014PhR...541...45C}) and here we assume $\tau=1$ at $\lambda_{\rm rest}=200$\micron. The dust temperature is set as a free parameter in the model and we extract our adopted values from the best fit FIR/mm SED. Assuming $z_{\rm PDF}$, $\lambda_{\rm peak}$ and $T_{\rm dust}$ are 83$^{+9}_{-8}$\,\micron\ and 60$^{+8}_{-7}$\,K for MORA-5 and 106$^{+19}_{-20}$\,\micron\ and 43$^{+14}_{-9}$\,K for MORA-9.


\subsection{Dust Mass}
\label{subsec:dust_mass}
We directly infer dust masses from the associated 2\,mm dust continuum detections on the Rayleigh-Jeans tail of blackbody emission (at $\lambda_{\rm rest}\gtrsim300\,\micron$ where dust emission is likely to be optically thin) following the framework of \cite{Scoville2016ApJ...820...83S}. The dust mass is proportional to the mass-weighted dust temperature (set to 25\,K, representative of the bulk of dust dispersed throughout the entire galaxy) and observed flux density at $\nu_{\rm obs}$. Given the potentially high-redshift solutions, we assume cosmic microwave background (CMB) heating of the dust in these sources is non-negligible, so we employ Equation 1 from \cite{Casey2019ApJ88755C} for our dust mass calculation as it incorporates a correction factor for suppressed flux density against the CMB background. We estimate dust masses of (3.6$\pm0.5)\times10^9$\,M$_\odot$ for MORA-5 and (2.2$\pm0.4)\times10^9$\,M$_\odot$ for MORA-9 with the associated errors propagated from the measurement error of the 2\,mm fluxes. The dust-to-stellar mass ratios ($M_{\rm dust}/M_\star$) for our sources is $0.024\pm0.018$ and $0.054\pm0.031$. When compared to a recent compilation of $M_{\rm dust}/M_\star$ for DSFGs \citep{Donevski2020A&A...644A.144D}, we find that both MORA-5 and MORA-9 lie above the determined scaling relation at their respective redshifts, which is identified to rise up to $z\sim2$ before a mild decline/flattening at $z\gtrsim2$ is observed. Contrary to low redshift galaxies, we find our higher $M_{\rm dust}/M_\star$ values to be unsurprising considering the high dust content is assumed to be primarily responsible for their defining features as OIR-dark and IR luminous sources. 



\subsection{Gas Mass}
Functionally, our gas mass estimates are determined in the same manner as dust mass as they are both inferred from the 2\,mm data with gas mass being scaled via a dust-to-gas ratio. We adopt a a CO-to-H$_2$ conversion factor of $\alpha_{\rm CO}=6.5$\,M$_\odot$\,(K\,km\,s$^{-1}$\,pc$^2$)$^{-1}$, accounting for the mass of both H$_2$ and He gas. This follows the methodology of \cite{Scoville2016ApJ...820...83S} and Equation 3 of \cite{Casey2019ApJ88755C} and accounts for CMB heating. 
This method makes use of the empirically calibrated conversion factor from 850\,\micron\ luminosity to interstellar medium (ISM) mass, $\alpha_{\rm 850}=(6.7\pm1.7)\times10^{19}$\,erg\,s$^{-1}$\,Hz$^{-1}$\,M$_\odot^{-1}$, as determined in \cite{Scoville2016ApJ...820...83S}, and intrinsic to this calculation is the assumed CO-to-H$_2$ conversion factor. We adopt the same mass-weighted dust temperature of 25\,K as we did in Section \ref{subsec:dust_mass}. Ultimately, we find $M_{\rm gas}=(2.6\pm0.4)\times10^{11}$M$_\odot$ for MORA-5 and $M_{\rm gas}=(1.6\pm0.3)\times10^{11}$M$_\odot$ for MORA-9.


\section{Discussion}
\label{sec:discussion}

\begin{figure}
    \centering
    \includegraphics[width=\columnwidth]{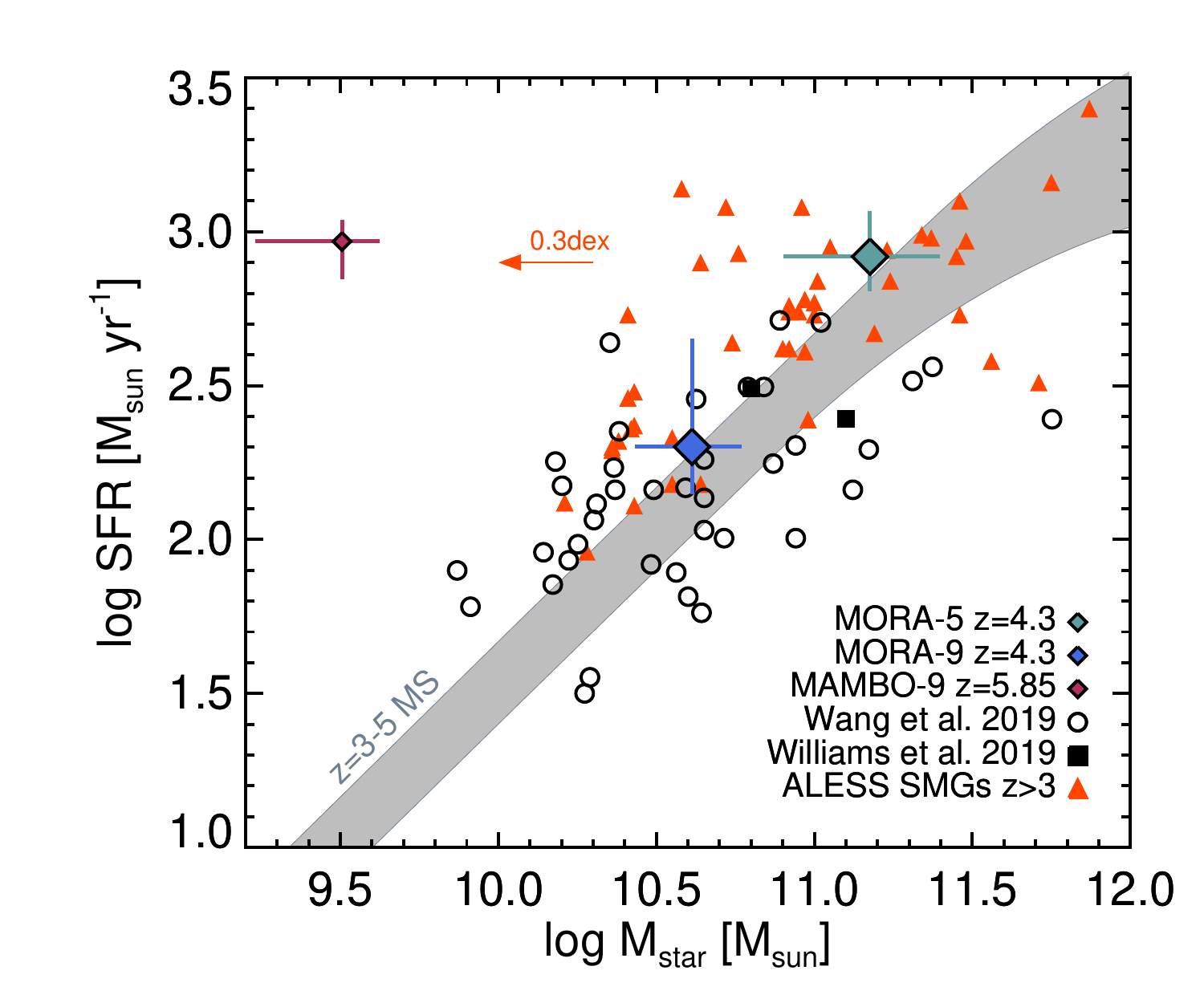}
    \caption{SFR versus M$_\star$ for MORA-5 (teal diamond) and MORA-9 (blue diamond) given their $z_{\rm PDF}$ photometric redshifts. MORA-4 (a.k.a. MAMBO-9) is shown for comparison (small maroon diamond) as one of the other OIR-dark DSFGs detected in the MORA survey. MORA-3 (a.k.a. AzTEC-2) is not depicted here due to its lack of stellar mass estimate -- a result of its IRAC photometry severely blending with foreground sources. We plot our results on top of H-band dropouts from \cite{Wang2019Nature} (open circles) and 3\,mm detected OIR-dark DSFGs \citep{Williams2019ApJ...884..154W} (filled squares) at similar redshifts from the literature. We also show $z>3$ ALESS DSFGs \citep{DaCunha2015ApJ806110D} (orange triangles) and the main sequence of star-formation at $z=3-5$ \cite{Schreiber2015A&A...575A..74S} (gray region). The length of the orange arrow depicts the 0.3\,dex offset in M$_\star$ applied to the ALESS sample in \cite{Wang2019Nature} due to the different methods used to calculate stellar masses.}
    \label{fig:sfr_mstar}
\end{figure}

\subsection{OIR-dark Galaxies: A Subsample of the SMG/DSFG Population}
\label{subsec:OIRdark Overview}
We suspect OIR-dark DSFGs make up part of the high-$z$ tail of the population of dust-obscured submillimeter galaxies \citep[SMGs;][]{Hughes1998Natur.394..241H,Blain2002PhR...369..111B,Casey2014PhR...541...45C}. Canonically selected at 850\,\micron\ with $S_{850}\gtrsim2$\,mJy, SMGs are the most well-studied DSFGs with over two decades of work dedicated to understanding their redshift distributions \citep{Chapman2005ApJ...622..772C,Zavala2014MNRAS.443.2384Z,Brisbin2017A&A...608A..15B}, number counts \citep{Scott2002MNRAS.331..817S}, spatial density \citep{Aravena2010ApJ...708L..36A,Smolcic2017A&A...597A...4S}, and intrinsic physical properties \citep{Miettinen2017A&A...606A..17M,Dudzevi2020MNRAS.494.3828D}. SMGs on average have reported stellar and gas masses $\sim10^{11}\,M_{\odot}$, SFRs in the range of 100--1000\,$M_\odot$\,yr$^{-1}$, and have often been cited as the progenitors of massive quiescent galaxies, e.g., ellipticals in the centers of local galaxy groups and clusters \citep{Hainline2011ApJ74096H}. Similarly, $z\gtrsim3$ SMGs have been suggested to be the parent source for intermediate redshift ($z\sim2$) compact quiescent galaxies \citep{Toft2014ApJ...782...68T}. Moving forward, we group SMGs under the more generalized term of DSFGs for simplicity, which encompasses all galaxies at high-$z$ originally selected at FIR or submillimeter wavelengths. The OIR-dark galaxies discussed in this work are part of this population, having also been first detected at (sub)mm wavelengths and exhibiting similarly broad ranges of stellar masses and star-formation activity.

The majority of known DSFGs at $z>4$ are almost all extreme star-forming systems (forming upwards of 1000 M$_\odot\,$yr$^{-1}$ and with stellar masses exceeding $10^{11}$\,M$_\odot\,$) as they are predominantly detected by {\it Herschel}. As with early DSFG observations, initial OIR-dark DSFG discoveries were of the most extreme, rare, starbursts \citep{Walter2012Natur.486..233W}, and not necessarily termed OIR-dark. Gas/dust-rich, optically undetected DSFGs have also been discovered as companion systems to (sub)mm bright high redshift quasars \citep{Fogasy2017A&A...597A.123F,Fogasy2020MNRAS.493.3744F}. Now, with the advent of longer wavelength interferometric surveys covering wider areas, independent of single dish studies, more moderate systems are being readily discovered and are believed to make up the bulk of the population of these highly obscured galaxies \citep{Fujimoto2016ApJS..222....1F,Oteo2016ApJ...822...36O,Franco2018A&A...620A.152F,Gonzalez-Lopez2020ApJ...897...91G,Zavala2021ApJ...909..165Z}. Interestingly, \cite{Valentino2020ApJ88993V} suggest galaxies with lower (sub)mm fluxes ($S_{\rm 850}<3.5$\,mJy) better reproduce the $M_\star$ and SFR distributions expected for the progenitors of high-$z$ quiescent galaxies, further supporting our interest in uncovering such systems. The extremely dust-obscured sources (undetected in the deepest $Ks$-band surveys) are proposed to make up 20\% of all DSFGs according to \cite{Dudzevi2020MNRAS.494.3828D}, which is in line with our results from the MORA survey as 23\% of the detected DSFGs are OIR-dark. 

\cite{Wang2019Nature} reported 39 OIR-dark DSFGs (selected as H-band dropouts) at $z_{\rm phot}>3$. They were detected via ALMA at observed-frame 870\,\micron\ and determined to be DSFGs via their robust millimeter detections. The median stellar mass of this population is $M_\star\sim10^{10.6}$M$_\odot$ with a characteristic IR luminosity of $L_{\rm IR(8-1000\micron)}=(2.2\pm0.3)\times10^{12}$L$_\odot$ and SFRs $\sim200$\,M$_\odot$\,yr$^{-1}$ as determined via stacking. Similarly, \cite{Williams2019ApJ...884..154W} reported serendipitous detections of two previously unknown $z>3$ sources found in a small 8\,arcmin$^2$ 3\,mm survey in the COSMOS field, one of which being an ``ALMA-only'' source (named $3MM-1$) with $z_{\rm phot}=5.5^{+1.2}_{-1.1}$ and OIR counterparts only detected ($\sim3\,\sigma$) in deep, stacked IRAC 3.6\,\micron\ and 4.5\,\micron\ observations (see also \citealt{Zavala2021RNAAS...5...15Z} which reports a tentative CO(6-5) line indicative of a $z=5.857$ redshift solution). Both sources have SFRs akin to ``normal" main sequence (MS) galaxies (a few 100\,M$_\odot$\,yr$^{-1}$) and are in line with the H-band dropouts on the MS. OIR-dark DSFGs have also been identified starting from deep radio surveys. For example, \cite{Talia2021ApJ...909...23T} present 197 VLA sources ($S_{\rm 3GHz}>12.65$\,\uJy\,beam$^{-1}$) which do not have a COSMOS2015 counterpart, though we note MORA-5 and MORA-9 would not be included given this selection criteria.

Based on our current observations and SFR calculations, MORA-9 appears to belong to the moderately star-forming subsample of OIR-dark DSFGs, while MORA-5 displays physical characteristics consistent with other rare, extreme OIR-dark starbursts. 
Figure \ref{fig:sfr_mstar} shows SFR versus M$_\star$ of our results of MORA-5 and MORA-9 (blue and teal diamonds) along with MORA-4, a.k.a. MAMBO-9, a spectroscopically confirmed OIR-dark DSFG (see \citealt{Casey2019ApJ88755C}). The H-band dropouts (open circles), 3\,mm detected OIR-dark DSFGs (filled squares), and $z>3$ DSFGs (orange triangles) from the ALESS survey \citep{DaCunha2015ApJ806110D} are shown for comparison. Stellar masses of the ALESS sample are reduced by 0.3\,dex in \cite{Wang2019Nature} (depicted by the orange arrow) to account for different assumptions in their respective mass estimations. Since stellar masses for the MORA sources are estimated in the same manner as the ALESS sample (through {\tt MAGPHYS}), we do not apply this offset in our figure. To compare apples to apples, the H-band dropout sample should have a factor of $\sim2$ higher mass than shown. These results are all plotted on top of the MS of galaxies at $z=3-5$ (filled gray region) as parameterized in \cite{Schreiber2015A&A...575A..74S}.

\subsection{The Dynamic Range of OIR-dark DSFGs}
\label{subsec:dynamic range of OIR-dark DSFGs}
This work, along with those mentioned in Section \ref{subsec:OIRdark Overview}, illustrates the heterogeneous nature of OIR-dark DSFGs. The MORA survey, with its relatively large coverage area (184\,arcmin$^{-2}$ in total) for an interferometric survey, has shown to be effective at detecting both rarer, bright OIR-dark DSFGs with high submillimeter fluxes as well as fainter sources -- a unique aspect of the survey design. We defined OIR-dark to simply include any galaxy (likely a DSFG) lacking detections at wavelengths shortward of 2.2\,\micron, and it is clear this encompasses a broad range of star-formation and stellar mass. 

While we lack some clarity on why some DSFGs are OIR-dark and others are not, it seems likely that a combination of factors are at play. Perhaps the most evident component contributing to the OIR-dark nature of DSFGs is the high redshifts of these sources compared to other DSFGs with comparable (sub)mm fluxes across $S_{\rm 870}=3-10$\,mJy. \cite{Smail2021MNRAS.502.3426S} finds that $K-faint$ ($K$-band magnitude\,$>25.3$) DSFGs from the AS2UDS survey \citep{Dudzevi2020MNRAS.494.3828D} have a significantly higher median photometric redshift compared to their brighter $K$-detected sample: $z=3.44\pm0.06$ versus $z=2.36\pm0.11$, respectively. This work also shows a higher median dust attenuation ($A_V=5.2\pm0.4$) for the $K-faint$ population compared to their redshift-matched control sample ($A_V=2.9\pm0.1$), which is attributed to the smaller dust continuum sizes observed (i.e. higher specific SFR, $\Sigma$SFR). Finally, \cite{Smail2021MNRAS.502.3426S} point to the mixture and relative scales of the obscured versus unobscured components, suggesting OIR-dark DSFGs may exhibit an absence of spatially extended, less obscured star-formation. The lower $A_V$ value for MORA-9 and its associated $K$-band detection suggests it has a stellar component which is geometrically distinct from the dust emission whereas those same components for MORA-5 may be coupled.  


\subsection{Do OIR-dark DSFGs Host Active Galactic Nuclei?}
We searched for X-ray emission in the \textit{Chandra} COSMOS Legacy data to infer the presence of AGN in MORA-5 and MORA-9, but neither source is detected. However, given the suspected high redshifts of these sources and the depth of the X-ray observations, the current data cannot rule out the presence of an AGN.

The infrared to 1.4\,GHz radio luminosity ratio, \qir, quantifies the far-IR/radio correlation (FRC). Various works have looked to determine if a redshift evolution of \qir\ as well as how it may differ in star-forming galaxies (SFGs) compared to AGN \citep{Magnelli2012A&A...539A.155M,Delhaize2017A&A...602A...4D}. Assuming a synchrotron slope of $\alpha=-0.8$ \citep{Condon1992ARA&A..30..575C}, the marginal ($3\sigma$) 3\,GHz detection of MORA-5 implies a lower limit of \qir\,$>1.9$ which is in agreement with what is predicted by the evolution of \qir\ for a $z=4.3$ SFG (\qir$=2.1$; \citealt{Delhaize2017A&A...602A...4D}). Examining the radio emission of MORA-9 produces similar results and we find a lower limit of \qir\,$>1.6$. 

When examined separately, AGN are found to have lower \qir\ values overall and at this redshift a galaxy with a luminous AGN could be expected to have \qir$<1.5$ given the evolution of the FRC found by \cite{Delhaize2017A&A...602A...4D}. Even with a marginal radio detection in MORA-5, the uncertainty in the source's redshift makes it impossible to conclude whether or not an AGN is present; for example, it may sit at lower redshift with radio emission from star-formation or at higher redshift with a radio loud AGN. Confirming the redshifts of the two sources via an emission line search will be critical in determining whether or not they host luminous AGN.

\subsection{Potential Evolutionary Tracks}
The discovery of OIR-dark DSFGs at $z>3$ has now been evolutionarily linked to the recent detections of high redshift (up to $z\sim4$) massive (M$_\star\sim10^{11}\,$M$_\odot$) quiescent galaxies \citep{Spitler2014ApJ,Glazebrook2017Natur,Schreiber2018bA&A,Valentino2020ApJ88993V,Stevans2021arXiv210314690S}. 
The prevalence of OIR-dark galaxies in the early Universe has large implications for the cosmic star-formation rate density (SFRD), stellar mass function (SMF), and theoretical models which have broadly excluded this population of galaxies due to their poor observational constraints \citep{Dudzevi2020MNRAS.494.3828D}.
These discoveries also drastically compress the formation and quenching timescales of quiescent galaxies to just a few billion years and raises questions regarding the galaxies they evolved from. While a handful of DSFGs have been discovered out to $z\sim7$ \citep{Marrone2018Natur.553...51M}, they represent rare systems and are not thought to make up the bulk of the population of DSFGs at $z>3$. Ultimately, the number density of massive high-redshift galaxies required to reproduce the observed population of quiescent galaxies at $z=2-4$ is not supported by the summation of $z>4$ massive LBGs and known extreme DSFGs alone \citep{Straatman2014ApJ...783L..14S}. To assess the viability of OIR-dark DSFGs as the progenitors of high-$z$ massive quiescent galaxies, we follow arguments based on number density and stellar mass evolution. 

\subsubsection{Space Density of OIR-dark DSFGs}
In total, four OIR-dark DSFGs were detected in the MORA survey. Along with MORA-5 and MORA-9, two additional sources which already have spectroscopic confirmations, MORA-3 (a.k.a. AzTEC-2; \citealt{Jimenez-Andrade2020ApJ...890..171J}) and MORA-4 (a.k.a. MAMBO-9; \citealt{Casey2019ApJ88755C}), are folded into our source density calculations. We rely on these spectroscopic redshifts (MORA-3: $z=4.63$ and MORA-4: $z=5.85$) to aid in our estimate of the upper bound of the selection volume, ultimately adopting $z=6$. Nominally, the maximum redshift of a 2\,mm-detected DSFG could be much higher given the negative K-correction, even after accounting for CMB heating. Realistically though, we know sources are extremely rare at $z>6$ based on current surveys and our understanding of the IR luminosity function \citep{Zavala2021ApJ...909..165Z}. We then set the lower bound to $z=4$, given our photometric redshifts for MORA-5 and MORA-9. Examining the contiguous 156\,arcmin$^2$ area where all four sources were detected across $4<z<6$ results in a number density of $0.03$\,arcmin$^{-2}$ and comoving volume density of $n\sim(5\pm2)\times10^{-6}$\,Mpc$^{-3}$. The uncertainty on the volume density is generated by running MC trials to create the distribution of expected sources given both cosmic variance (35\% between $3<z<6.5$ as determined in \citealt{Zavala2021ApJ...909..165Z}) and Poisson noise. These distributions are then summed and the resulting $1\,\sigma$ errors are reported. The area of the survey is sufficiently large, despite the small sample size, for this first order estimate given that cosmic variance only begins to dominate beyond $z>6.5$ \citep{Zavala2021ApJ...909..165Z}. A larger survey area, however, would of course improve our determination of number density. 

This result is $\sim4\times$ lower than the total H-band dropout space density reported in \cite{Wang2019Nature} ($n\sim2\times10^{-5}$\,Mpc$^{-3}$), which is said to be in agreement with that of massive quiescent galaxies found at $z\sim3$ \citep{Straatman2014ApJ...783L..14S}. We note that our sample is ALMA-selected, whereas $\sim37$\% of the H-band dropouts are ALMA-undetected. Limiting the calculation to H-band dropouts with ALMA detections lowers the volume density of the \cite{Wang2019Nature} sample by a factor of 2. Furthermore, if we look only at the twenty $4<z<6$ sources in the sample, the density drops down to $n\sim6\times10^{-6}$\,Mpc$^{-3}$ and falls in line with what we observe in MORA across this redshift range.

Whether or not a source is observed to be OIR-dark, all DSFGs are expected to quench and evolve into passive elliptical galaxies. Indeed, being OIR-dark is likely very dependent on redshift and geometry as discussed in Section \ref{subsec:dynamic range of OIR-dark DSFGs}. With this in mind, we also calculate the volume density of all 2\,mm-detected DSFGs in the MORA sample from $3<z<6$ and find $n\sim1\times10^{-5}$\,Mpc$^{-3}$. While it is reassuring to see agreement across different samples, the measurements of spatial density presented in the literature as well as this work both suffer from uncertainties in the co-moving volume due to the reliance on photometric redshifts, highlighting the need for future spectroscopic follow-up of high-$z$ DSFGs.

\subsubsection{Stellar Mass Growth}
 Both MORA-5, $M_\star=(1.5^{+1.0}_{-0.7})\times10^{11}$\,M$_\odot$, and MORA-9, $M_\star=(4.1^{+1.8}_{-1.4})\times10^{10}$\,M$_\odot$, have stellar masses already consistent with $z\sim3$ massive quiescent galaxies \citep{Schreiber2018bA&A}, albeit on the lower mass end for MORA-9. High-$z$ DSFGs are commonly thought to undergo a short ($\sim50-150$\,Myr), bursty period of star-formation mediated by strong galactic feedback \citep{MihosHernquist1996ApJ...464..641M,Cox2008MNRAS.384..386C,Toft2014ApJ...782...68T,Swinbank2014MNRAS.438.1267S,Aravena2016MNRAS.457.4406A, Spilker2020ApJ...905...86S}, potentially triggered by an initial gas-rich major merger or interaction. Conversely, consistent gas infall may lead to significantly longer duty cycles of up to a gigayear \citep{Hayward2013MNRAS.428.2529H,Narayanan2015Natur.525..496N}. Based on the star-formation rates and gas masses estimated in Section \ref{sec:results}, MORA-5 and MORA-9 have star-forming gas depletion timescales of $\tau_{\rm depl}\sim310$\,Myr and $\tau_{\rm depl}\sim800$\,Myr, respectively. This is significantly longer than DSFGs at $z\sim2-3$. Assuming these gas depletion times, the stellar mass of MORA-5 increases to $\sim4\times10^{11}$\,M$_\odot$ by $z=3.6$ and the stellar mass of MORA-9 increases to $\sim2\times10^{11}$\,M$_\odot$ by $z=2.9$. These evolved masses are still in line with the mass range of the quiescent sample, however it should be noted that we cannot exclude the possibility of accretion refilling the gas reservoirs and extending the depletion timescales. \cite{Valentino2020ApJ88993V} (Figure 6 therein) illustrate the wide range in predicted number densities of quiescent galaxies with log($M_\star/M_\odot$)\,$>10.6$ from both observations and theory across $3\lesssim z \lesssim4$. This mass threshold is well aligned with the entire MORA sample for which we expect to produce galaxies with stellar masses in excess of log($M_\star/M_\odot$)\,$>10.6$ (\citealt{Casey2021arXiv211006930C}, Figure 13). For the COSMOS field specifically, the stellar mass function determined in 
 \cite{Davidzon2017A&A...605A..70D} produces a number density of $n\sim2\times10^{-6}$\,Mpc$^{-3}$, which then drops precipitously as the mass threshold increases. The lack of consensus in the number density of $z>3$ quiescent galaxies speaks to the need for future observations with JWST to detect and constrain this galaxy population. Luckily, this is a science goal for several of the approved Cycle 1 proposals and as such we eagerly await those results. 
 
If we consider the current stellar mass estimates alone in this closed box scenario, the MORA OIR-dark DSFGs are viable progenitors to high-$z$ quiescent galaxies. For now, our conjecture stops there until more accurate stellar masses and greater understanding of potential feedback and quenching mechanisms for these two OIR-dark DSFGs can be obtained. Future observations with JWST MIRI imaging would provide constraints on the rest-frame 1.6\,\micron\ stellar bump, thus elucidating the stellar masses of these heavily obscured systems.

\section{Conclusions}
We present photometric redshifts and physical characterization of MORA-5 and MORA-9, two OIR-dark DSFGs detected in the 2\,mm MORA survey. These sources, of interest for their potential to be the highest redshift objects in MORA, lacked secure detections in any photometric bands short of 2.2\,\micron, consistent with other OIR-dark systems reported in recent literature.

The photometric redshifts reported in this work, $z_{\rm PDF}=4.3^{+1.5}_{-1.3}$ for MORA-5 and $z_{\rm PDF}=4.3^{+1.3}_{-1.0}$ for MORA-9 are the result of combining the redshift distributions determined via OIR, FIR/mm, and energy balance SEDs. It is this $z_{\rm PDF}$ value which informs the physical properties derived for these two systems with the exception of stellar mass and $A_V$. These two properties are instead determined by \magphysphotz\ given its inclusion of IR, (sub)mm, and radio data which is especially beneficial for high-$z$, dust-obscured galaxies. 

Based on our current observations, MORA-5 is the more active of the two systems. Its high stellar mass, (1.5$^{+1.0}_{-0.7})\times10^{11}$\,M$_\odot$, and star-formation rate, ${\rm SFR}\approx830$\,M$_\odot$yr$^{-1}$, suggests it is part of a class of rarer, more extreme OIR-dark galaxies. MORA-9 is modest in comparison with (4.1$^{+1.8}_{-1.4})\times10^{10}$\,M$_\odot$ and ${\rm SFR}\approx200$\,M$_\odot$yr$^{-1}$. From our 2\,mm dust continuum observations we determine MORA-5 and MORA-9 to have gas masses of $M_{\rm gas}=(2.6\pm0.4)\times10^{11}$M$_\odot$ and $M_{\rm gas}=(1.6\pm0.3)\times10^{11}$M$_\odot$ and gas depletion timescales of $\tau_{\rm depl}\sim310$\,Myr and $\tau_{\rm depl}\sim800$\,Myr, respectively. Evolving the stellar masses according to our estimated gas depletion timescales results in both systems remaining consistent with the masses of known $z\sim3$ quiescent galaxies.

It is promising to see convergence towards $z>3$ solutions in all the redshift PDFs given the decision to observe at 2\,mm to effectively weed out $z\lesssim2.5$ interlopers thanks to the very negative K-correction at millimeter wavelengths. However, we still highlight the need for spectroscopic follow-up of OIR-dark DSFGs to confirm redshifts and allow for accurate calculations of their spatial density and contribution to the cosmic star-formation rate density (CSFRD) as a function of redshift. Secure redshifts are also essential for placing constraints on dust production mechanisms and build-up within the first billion years after the Big Bang. Redshift confirmation via spectral scan to search for CO transitions with ALMA could be achieved in less than eight hours for both sources combined, taking into account the required rms sensitivity determined by the $L_{\rm IR}$ -- $L_{\rm CO}$ relation \citep{Bothwell2013MNRAS.429.3047B,Greve2014ApJ...794..142G}. 

Given the uncertainties in redshift (and thus, SFR) and stellar masses of OIR-dark DSFGs, it is difficult to compare them directly with high-$z$ quiescent galaxies. Based on our current understanding of the space density of both populations and estimates of stellar mass growth, our work is in agreement with previous studies and supports the notion that OIR-dark DSFGs are the progenitors of $z=2-4$, massive, quiescent galaxies. Resolved (sub)mm observations providing effective radius measurements of OIR-dark DSFGs are necessary for determining an evolutionary link to compact quiescent galaxies in terms of their comparative spatial extents. They would also allow for measurements of the average dust column density (alluding to the interstellar medium conditions in OIR-dark DSFGs) as well as star-formation surface density. Future observations, either with JWST NIRSpec campaigns or emission line surveys (e.g. with ALMA, NOEMA, LMT) would establish accurate redshifts, while JWST NIRCam and MIRI imaging would constrain the stellar emission and offer insights into the morphologies of these systems. Finally, we will look to expand with wide-field 2\,mm imaging as an effective strategy to detect high-$z$, OIR-dark DSFGs. 

\section*{acknowledgments}
We thank the anonymous reviewer for their thoughtful suggestions which helped improve this manuscript. SMM thanks Isabella Cortzen, Seiji Fujimoto, and Francesco Valentino for their hospitality and helpful discussions while visiting the Cosmic DAWN Center. SMM also thanks her parents who provided a home, companionship, and a desk to write this paper during the lockdown brought on by the COVID-19 pandemic. Finally, SMM thanks the National Science Foundation for support through the Graduate Research Fellowship under Grant No. DGE-1610403, the NSF GROW travel grant, Georgios Magdis for being her site host at DAWN, and the NRAO SOS grant. Support for this work was also provided by NASA through the NASA Hubble Fellowship grant \#HST-HF2-51484 awarded by the Space Telescope Science Institute, which is operated by the Association of Universities for Research in Astronomy, Inc., for NASA, under contract NAS5-26555. CMC thanks the National Science Foundation for support through grants AST-1814034 and AST-2009577 and thanks RCSA for a Cottrell Scholar Award, sponsored by IF/THEN, an initiative of the Lyda Hill Philanthropies. GEM acknowledges  the Villum Fonden research grant 37440, "The Hidden Cosmos” and the Cosmic Dawn Center of Excellence funded by the Danish National Research Foundation under then grant No. 140. MA acknowledges support from FONDECYT grant 1211951, ``CONICYT + PCI + INSTITUTO MAX PLANCK DE ASTRONOMIA MPG190030'' and ``CONICYT+PCI+REDES 190194''. KK acknowledges support from the Knut and Alice Wallenberg Foundation. MT acknowledges the support from grant PRIN MIUR 2017. This paper makes use of the following ALMA data: ADS/JAO.ALMA\#2019.2.00143.S. ALMA is a partnership of ESO (representing its member states), NSF (USA) and NINS (Japan), together with NRC (Canada), MOST and ASIAA (Taiwan), and KASI (Republic of Korea), in cooperation with the Republic of Chile. The Joint ALMA Observatory is operated by ESO, AUI/NRAO and NAOJ. The National Radio Astronomy Observatory is a facility of the National Science Foundation operated under cooperative agreement by Associated Universities, Inc.

\software{CASA \citep[v5.6.1;][]{McMullin2007ASPC..376..127M}, EAZY \citep[v1.3;][]{Brammer2008ApJ686}, MAGPHYS \citep{DaCunha2008MNRAS3881595D}, MAGPHYS+photo-z \citep{Battisti2019ApJ...882...61B}, LePhare \citep{Arnouts2002MNRAS.329..355A,Ilbert2006A&A...457..841I}.}

\bibliographystyle{aasjournal}

\bibliography{alma2mm}

\end{document}